\documentclass[12pt]{article}
\usepackage{caption}
\usepackage{subcaption}
\usepackage{geometry}
\usepackage{titlesec}
\usepackage{multirow}
\usepackage[T1]{fontenc}
\usepackage[utf8]{inputenc}
\usepackage{authblk}
\usepackage{graphics}
\usepackage{makecell}
\usepackage{float}
\usepackage{bigints} 
\usepackage[dvipsnames]{xcolor}
\providecommand{\keywords}[1]{\textbf{Keywords: } #1}
\titleformat{\section}{\centering\large\scshape}{\thesection}{1em}{}
\titleformat{\subsection}{\centering\normalsize\scshape}{\thesubsection}{1em}{}

\usepackage{hyperref}
\hypersetup{
	colorlinks=true,
	linkcolor=blue,
	filecolor=blue,      
	urlcolor=blue,
	citecolor=blue
}

\usepackage{natbib}
\usepackage{graphicx} 
\graphicspath{{art/}}

\usepackage{tikz}
\usetikzlibrary{fit,positioning,arrows,automata,calc}
\tikzset{
	main/.style={circle, minimum size = 5mm, thick, draw =black!80, node distance = 10mm},
	connect/.style={-latex, thick},
	box/.style={rectangle, draw=black!100}
}
\usetikzlibrary{decorations.pathreplacing}

\usepackage[T1]{fontenc}
\usepackage{lmodern}

\usepackage{setspace}
\usepackage{xr}
\usepackage{color}
\usepackage{amsmath, amssymb}
\usepackage{amsfonts,dsfont,mathrsfs,mathtools}
\usepackage{booktabs,multirow,array}
\usepackage{bm}
\usepackage{arydshln}
\usepackage[vlined,linesnumbered,ruled,resetcount]{algorithm2e}

\newcommand{\R}{\mathds{R}}

\newcommand{\Law}{\mathcal{L}}

\def\simiid{\stackrel{\mbox{\scriptsize{iid}}}{\sim}}

\allowdisplaybreaks[3]

\newcommand{\pkg}[1]{{\fontseries{m}\fontseries{b}\selectfont #1}}
\let\proglang=\textsf
\newcommand{\fct}[1]{\texttt{#1}}
\newcommand{\odv}[2]{\frac{d #1}{d #2}}

\newcommand{\code}[1]{\texttt{\detokenize{#1}}}

\usepackage{fancyvrb} 
\usepackage{xcolor}   

\newenvironment{CodeChunk}{}{}

\DefineVerbatimEnvironment{CodeInput}{Verbatim}{fontshape=sl}

\DefineVerbatimEnvironment{CodeOutput}{Verbatim}{}

\begin{document}

\def\spacingset#1{\renewcommand{\baselinestretch}%
{#1}\small\normalsize} \spacingset{1}


\title{\bfseries\LARGE{BayesChange: an \proglang{R} package for Bayesian Change Point Analysis}}
 
\author[1]{Luca Danese}
\author[1]{Riccardo Corradin}
\author[1]{Andrea Ongaro}
\affil[1]{\normalsize{Department of Economics, Management and Statistics, University of Milano-Bicocca, Milano, Italy}}

\date{}

\maketitle

\bigskip
\begin{abstract}
	
\noindent
We introduce \pkg{BayesChange}, a computationally efficient \proglang{R} package, built on \proglang{C++}, for Bayesian change point detection and clustering of observations sharing common change points. While many \proglang{R} packages exist for change point analysis, \pkg{BayesChange} offers methods not currently available elsewhere. The core functions are implemented in \proglang{C++} to ensures computational efficiency, while an \proglang{R} user interface simplifies the package usage. The \pkg{BayesChange} package includes two \proglang{R} wrappers that integrate the \proglang{C++} backend functions, along with \proglang{S3} methods for summarizing the results. We present the theory beyond each method, the algorithms for posterior simulation and we illustrate the package's usage through synthetic examples.
\vspace{12pt}

\noindent\keywords{change point detection, Bayesian statistics, C++, model based clustering, \proglang{R}}
\end{abstract}
  
\section{Introduction}

Detecting structural changes in time-dependent data is a fundamental topic in modern statistics. For example, one might be interested in finding structural changes in financial time series, the time instants where the outbreak of a pandemic changes its behavior, or changes in the functioning of a specific machine. In statistical analysis, these structural changes are referred to as \textit{change points}, and they occur whenever a time-dependent data generating process changes its behavior, usually meaning that it changes the local value of its parameters. Seminal works on change point detection are based on hypothesis testing, where the null hypothesis is the presence or absence of a single change point. The works of \cite{Page54,Pag57} and \cite{Cher64} are the first contributions having a frequentist and a Bayesian approach, respectively, following this direction. Among all methods, some of the Bayesian extensions provide significant flexibility and key modeling features, as they do not require one to specify the number and the position of change points and allow for quantification of uncertainty in the estimates. \pkg{BayesChange} implements exclusively Bayesian methods based on Product Partition Models (PPM). This class of models was first introduced by \cite{Har90}, and then applied to the problem of change point detection by \cite{bar92,bar93}. Later, PPM-based approaches to change point detection were proposed by \cite{Los03}, \cite{Qui03} and \cite{Fue10}. More recent developments include contributions by \cite{MM2014} and \cite{corradin_2022}, whose methods are implemented in \pkg{BayesChange}.


\pkg{BayesChange} is written in \proglang{C++} with an \proglang{R} \citep{R-base} interface, providing Bayesian methods for change point analysis. The package offers three main contributions:
\begin{enumerate}
    \item Implementation of the methods by \cite{MM2014} and \cite{corradin_2022} for change point detection in time series and epidemic diffusions associated with susceptibles-infected-removed (SIR) models.
    \item Implementation of the method by \cite{corradin2024} for clustering time-dependent data with common change points.
    \item \proglang{S3} methods for graphical visualization and estimation of results from both change points detection and clustering. 
\end{enumerate}

While several \proglang{R} packages provide functions that perform change point detection, \pkg{BayesChange} not only implements methods that are currently unavailable elsewhere, but also incorporates recent and more flexible Bayesian models proposed in the latest literature. The methods by \cite{MM2014} and \cite{corradin_2022}, allow change point detection in univariate and multivariate time series without requiring the user to pre-specify the number of change points. Moreover, thanks to the Bayesian framework, they naturally provide uncertainty quantification for both the number and locations of change points through the posterior distribution. The model proposed by \cite{corradin2024}, on the other hand, addresses a problem that no other models currently solve: clustering time-dependent data based solely on shared change point locations as the unique criterion for commonality.


Furthermore, methods in \pkg{BayesChange} are all based on the PPM theory. According to this framework, realizations are associated with a sequence of latent parameters, possibly with ties. If the parameter changes value, then a change point occurs. Consecutive realizations sharing the same parameter form a block, resulting in a latent order induced by the change points. The probability of this latent order is proportional to the product of functional of the block sizes. The \pkg{bcp} package by \cite{Erdman2007} implements the change points detection procedure by \cite{bar93}, but unfortunately it has been recently archived from the Comprehensive \proglang{R} Archive Network. The function \fct{cpp\_ppm} of \pkg{ppmSuite} \citep{ppmSuite} performs change points detection for univariate and multivariate time series, following the methodology of \cite{Quinlan2024}, assuming a correlation structure between change points. Finally, package \pkg{HDcpDetect} \citep{HDcpDetect} is not based on product partition models, but implements two Bayesian methods for change point detection. Specifically, a bisection algorithm strategy and an approach based on testing a large number of possible change points structure, selecting the one that best fits the data. 


Even though our main interest is on the Bayesian framework, we mention some packages that perform change point detection with a frequentist approach. The \pkg{decp} package \citep{decp}, recently made available on CRAN, provides two methods to infer changes on the covariance structure of multivariate time series, the \pkg{ecp} library \citep{ecp-article} includes two approaches for nonparametric change point detection on multivariate data and the \pkg{cpm} package \citep{cpm-article} perform both detection and prediction of change points on univariate time series. There are also packages that do not perform detection of change points, but take into account changes in the data when performing other analysis, like the \pkg{mixtools} package by \cite{Benaglia2009},  that perform regression modeling possibly accommodating for change points in the predictors. Finally, since \pkg{BayesChange} includes also a method for clustering epidemic diffusions arising from SIR models, that share the same changes, we also mention \pkg{CPsurv} \citep{CPsurv} which detects change points on survival data using a nonparametric approach.  


The paper is structured as follows. Section~\ref{sec:models} introduces the methodologies implemented in \pkg{BayesChange}. Section~\ref{sec:pkg_structure} provides details on the functions and the \proglang{S3} methods of the package. Finally, Section~\ref{sec:illustrations} shows how to use \pkg{BayesChange} in practice, trough synthetic data examples. Additional materials, such as details on the algorithms and the data generating processes, are deferred to the appendix. 

\section{Models and posterior estimates} \label{sec:models}

In this section, we introduce the theory behind the methods included in \pkg{BayesChange}. \pkg{BayesChange} includes a total of six different methodologies, to carry out different types of change point analysis. Three of these perform change points detection respectively on univariate time series, multivariate time series and on a single epidemic diffusion arising from an SIR model. The other three methods perform model-based clustering of samples of the previous, on the base of common change points. 

In general, we assume that a single observation is a vector $\bm{y}_i = \{ y_{i1}, \dots, y_{iT} \}$ of $T$ time realizations. If the observation is a time series, then $y_{it} \in \R^d$, with $d = 1$ for the univariate case and $d > 1$ for the multivariate one. For the epidemic diffusion arising from SIR models, $y_{it} \in \mathbb{N}^+$, as $y_{it}$ denotes to the number of new infected individuals in population $i$ on day $t$. The data generating distribution is a function $f$ that depends on a parameter $\theta_{it}$, describing the local behavior, that is specific for each observation at each time,
\begin{equation} \label{eq:realisation}
    \begin{split}
        y_{it} \sim f(\cdot \mid \theta_{it}).
    \end{split}
\end{equation}
For example, if realizations are generated by a Gaussian distribution, and we assume that both the mean and the variance are time dependent, then $\theta_{it} = (\mu_{it}, \sigma^2_{it})$. A change point for the time series occurs whenever $\theta_{it}$ changes in time. Hence, if $\theta_{it-1} \neq \theta_{it}$ then a change point occurs at time $t$ for $\bm{y}_i$. Change points induce a latent order of the realizations in $m_i$ blocks, here denoted with  $\rho_i = \{A_{i1}, \dots, A_{im_i}\}$. A latent order is nothing but a constrained latent partition, satisfying the following conditions: (i) $A_{ij} \cap A_{i\ell} = \emptyset$, for $j \neq \ell$, (ii) $A_{i1} \cup A_{i2} \cup \cdots \cup A_{im_i} = \{1, \dots, T\}$ and (iii) if $r \in A_{ij}$ and $s \in A_{i\ell}$, with $j < \ell$, then $r < s$. The generic $j$th block of the $i$th latent order is associated with an unique value of the latent parameter $\bm \theta_{ij}^*$. Two observations $y_{it}$ and $y_{i\ell}$ belong to the generic $j$th block if they share the same value of such parameter, $\theta_{it} = \theta_{i\ell} = \theta_{ij}^*$. Detecting change points in $\bm{y}_i$ means finding a point estimate for $\rho_i$, while clustering time dependent data with common change points means grouping together observations with the same latent orders. In the following subsections we present in details the two procedures. 

\subsection{Detecting change points on time series} \label{subsec:detect_cp}

In this package, methods for detecting change points on time series are based on the works by \cite{MM2014} and \cite{corradin_2022}. The former proposes a Bayesian nonparametric procedure to detect change points while the latter is a generalization of the first method to multivariate time series. In both cases, following \cite{MM2014}, the prior probability distribution of the latent order $\rho_i$ is devised by restricting the exchangeable partition probability function (eppf) of a Pitman-Yor process to the space of orders. Hence, the resulting distribution of the latent order equals
\begin{equation*}
    \begin{split}
        \Law(\rho_i) = \frac{T!}{m_i!} \frac{\prod_{j=1}^{m_i-1} (\delta + j\sigma) }{(\delta + 1)_{T-1}} \prod_{j=1}^{m_i} \frac{(1-\sigma)_{|A_{ij}|-1}}{|A_{ij}|},
    \end{split}
\end{equation*}
where $|A_{ij}|$ denotes the cardinality of the $j$th block of $\bm{y}_i$, i.e. the number of observations assigned to such block, and $\sigma \in (0,1)$ and $\delta \in (-\sigma, \infty)$ denotes the discount and strength parameters of the Pitman-Yor process, respectively. With a straightforward application of the Bayes' rule, the posterior probability of $\rho_i$ is then proportional to the product of the prior distribution $\Law(\rho_i)$ and the likelihood of $\rho_i$ given $\bm{y_i}$. 

We assume a markovian regime structure for $\bm{y}_i$. Specifically, if $d=1$ the distribution of $y_{it}$ is given by an univariate Ornstein-Uhlenbeck process,
\begin{equation*}
    \begin{split}
        y_{it} \mid \mu, \lambda, \phi \sim OU(\mu,\lambda,\phi), 
    \end{split}
\end{equation*}
where $\mu \in \mathbb{R}$ is the mean, $\lambda > 0$ the variance, and $0 < \phi < 1$ denotes the correlation between an observation at time $t$ and the one at time $t-1$. A priori, we assume a Normal-Gamma distribution for $(\mu, \lambda)$, with $\mu \mid \lambda \sim \text{N}(0,(c\lambda)^{-1})$ and $\lambda \sim \text{Ga}(a,b)$ for $\lambda$. Since we are not interested in the behavior of the time series, but only in the position of change points, $\mu$ and $\lambda$ are then integrated out of the likelihood, while $\phi$ is kept explicitly in the model. An explicit expression for the marginal likelihood is given by \cite{MM2014}. If $d > 1$, we set as distribution of $y_{it}$ a multivariate Ornstein-Uhlenbeck process, 
\begin{equation*}
    \begin{split}
        y_{it} \mid \bm{\mu}, \Lambda, \phi \sim OU(\bm{\mu},\Lambda,\phi), 
    \end{split}
\end{equation*}
where $\boldsymbol{\mu} \in \mathbb{R}^d$ is the mean vector, 
$\boldsymbol{\Lambda}$ is a $d \times d$ symmetric positive definite covariance matrix, and $0 < \phi < 1$ denotes the temporal correlation parameter. Here, $(\bm{\mu},\Lambda)$ are jointly distributed a priori as a Normal-Inverse Wishart distribution, with $\bm{\mu} \mid \Lambda \sim \text{N}(\bm{m}_0, k_0 \lambda)$ and $\Lambda \sim IW(\nu_0,S_0)$, where $\bm{m}_0 \in \R^d$, $k_0 > 0$, $\nu_0 > d-1$ and $S_0$ is a positive definite $d\times d$ matrix. Similarly to the univariate case, the parameters $\bm \mu$ and $\Lambda$ are integrated out of the likelihood. Explicit calculations can be found in \citet{corradin_2022}. 
In both univariate and multivariate cases, the law of the generic $i$-th observation is given by the product of the law of each block: 
\begin{equation*}
    \begin{split}
        \Law(\bm{y}_i \mid \rho_i, \bm{\theta}_i^*) = \prod_{j=1}^{m_i} \prod_{t_{ij}^-}^{t_{ij}^+} \Law(y_{it} \mid y_{it-1}, \theta_{ij}^*),
    \end{split}
\end{equation*}
where $t_{ij}^{-} = \min\{t : t \in A_{ij}\}$ and $t_{ij}^{+} = \max\{t : t \in A_{ij}\}$ are, respectively, the first and last time index of block $A_{ij}$.

\subsubsection{Posterior simulation}

To obtain a posterior estimate of the latent order $\rho_i$, both methods implement the same Markov Chain Monte Carlo (MCMC) algorithm based on a split-and-merge scheme, in the spirit of \citet{MM2014}. This procedure is specifically tailored for sampling from the posterior distribution of discrete objects such as latent orders or latent partitions. For seminal works on this topic, see \cite{greenrichardson_2001} and \cite{Jain2004ASM}. At each step of this procedure with probability $q$ a split is performed: a block is randomly chosen along with one observation that it belongs to. The block is then divided into two new blocks, with the chosen observation being the last realization of the first block and the following observation the first of the new block. With probability $1-q$ instead, a block is randomly chosen and its observations are merged with the observations of the following block. In both cases a new latent order is proposed where two blocks are merged together or an additional block is created. Once a new value for the latent order is proposed, the algorithm performs a Metropolis-Hastings step to accept or reject such value. Finally, if the number of blocks in the proposed order is greater than $1$, a shuffle step is performed in which observations of two adjacent blocks are rearranged while keeping the number of blocks unchanged. The output of the algorithm is a sequence of latent orders, one for each iteration, which represents a sample from the posterior distribution of $\rho_i$ given a sequence of real-valued time-dependent observations. 
In addition, the algorithm includes posterior sampling for the restricted eppf parameters, $\sigma$ and $\delta$, and for the parameter $\phi$. For parameters $\sigma$ and $\phi$ at each step a new value is proposed and evaluated with a Metropolis-Hastings procedure, while for $\delta$ at each step a value is extracted from its full conditional probability, whose form is known. Details about the steps of this procedure can be found in Section~\ref{app:algorithms} of the Appendix. 

The point estimate for the latent order is obtained by selecting from the posterior sample the order that minimizes a specific loss function, such as the \textit{Binder Loss function} \citep{Binder1978} or the \textit{Variation of Information} \citep{Wade2018}. For this purpose \pkg{BayesChange} also implements a method that selects the final order using a search algorithm called SALSO, introduced by \cite{dahl2022search}. 

Both methods for detecting change points were designed specifically for time series. However, \pkg{BayesChange} also provides a function to detect the change points on an epidemic diffusion arising from an SIR model. Here, the marginal distribution of the data term is given by evaluating the density function associated with such epidemic diffusion, where local parameters appearing in the likelihood are integrated numerically. More details are provided in Section~\ref{subsec:epi_data}.

\subsection{Clustering time-dependent data with common change points} \label{subsec:clust_data}

\pkg{BayesChange} also includes methods for clustering time series or survival functions that share common change points. These methods are based on a novel methodology introduced by \cite{corradin2024}, which clusters time-dependent data that share the same change point locations, without assuming any other commonalities. As with the change point detection model, the underlying algorithm is the same for both time series and survival functions, differing only in the distribution and the likelihood of the data. We first start by presenting the method in the context of time series, and later we discuss its extension to epidemiological data. 

Consider a set of time series $\mathcal{Y} = \{ \bm{y}_1, \dots,  \bm{y}_n\}$, where each realization of the time series is distributed according to Equation~\ref{eq:realisation}. The object of interest in this case is a random partition of $\{1, \dots, n\}$ in $k$ groups, here denoted with $\lambda = \{ B_1, \dots, B_k \}$, for which (i) $B_i \cap B_j = \emptyset$, for $i \neq j$, and (ii) $B_1 \cup \cdots \cup B_k = \{1, \dots, n\}$. Two generic observations $\bm{y}_i$ and $\bm{y}_j$ belong to the same group $B_l$ if they have the same change points, so if their latent orders $\rho_i$ and $\rho_j$ correspond. The sequence $ \mathcal{R}^* = \{\rho_{(1)}^\dagger, \dots, \rho_{(k)}^\dagger\}$ contains the unique latent orders associated with each block in $\lambda$. Thus, if $\bm{y}_i$ and $\bm{y}_j$ belong to the same block $B_l$, then $\rho_i = \rho_j = \rho_{(l)}^\dagger$. It is important to stress that the only commonality considered in this clustering method is the position of the change points, while the behavior of the time series is not taken into account as clustering criterion. Time series generated by distributions with different parameters but where changes happen at the same times belong to the same group. The model implemented in this method has the following hierarchical form  
\begin{equation}\label{mod:spec}
	\begin{split}
		\bm y_i \mid \rho_i, \bm \theta_i^* &\sim \prod_{j=1}^{m_i} \prod_{t = t_{ij}^-}^{t_{ij}^+} \Law(y_{it}\mid y_{it-1}, \theta_{ij}^*),\quad i = 1, \dots, n,\\
		\rho_i \mid \tilde p(\rho)&\simiid \tilde p(\rho) = \sum_{r=1}^{2^{T-1}}\pi_r \delta_{\Tilde{\rho}_r}(\rho), \quad i = 1, \dots, n,\\
		(\pi_1,\dots,\pi_{2^{T-1}}) &\sim \textsc{Dir}(\alpha_1, \dots, \alpha_{2^{T-1}}),\\
		\theta_{i,j}^*&\simiid P_0(\theta),\quad j = 1, \dots, m_i, \; i = 1, \dots, n,
	\end{split}
\end{equation}
where $t_{ij}^- = \min\left\{t: t \in A_{ij} \right\}$ and $t_{ij}^+ = \max\left\{t: t \in A_{ij} \right\}$. The latent order $\rho_i$ is sampled from a finite discrete mixture, where the weights are distributed according to a Dirichlet and the atoms are all possible $2^{T-1}$ orders of $\bm{y}_i$. The parameters of the Dirichlet distribution are assumed to be all equal $\alpha_1 = \dots = \alpha_{2^{T-1}} = \alpha$, which leads to a symmetric prior. The distribution $\Law(y_{it}\mid y_{it-1}, \theta_{ij}^*)$ depends on the nature of the data we consider, if we consider univariate and multivariate time series, we use the same approach based on the Ornstein-Uhlenbeck process presented in Section~\ref{subsec:detect_cp}. When we model epidemiological diffusion data, the distribution $\Law(y_{i,t}\mid y_{i,t-1}, \theta_{i,j}^*)$ is the likelihood of a survival function of an SIR model. 

\subsubsection{Posterior simulation}

To obtain a posterior estimate of $\lambda$ we implement the algorithm presented in \cite{corradin2024}. The procedure is still based on a split and merge procedure, but with the addition of a second level of sampling. Here, at each iteration both the latent partition of the observations $\lambda$ and the unique latent orders in $\mathcal{R}^*$ are updated. At each step two observations are randomly chosen, if they belong to the same group a random split of this group is proposed. Alternatively, if they belong to different groups the two groups are merged in a single group. Then, we perform a Metropolis-Hastings step to accept the proposed partition. In order to evaluate the acceptance rate, we need to assign latent orders to the new blocks of the proposed latent partition. These latent orders, here denoted by $\rho$, are obtained using an instrumental proposal distribution of the following form
\begin{equation} \label{eq:proposal}
    \begin{split}
        \psi(\rho \mid \mathcal{Y}) = \sum_{i=1}^{n}\frac{1}{n}\Law(\rho \mid \bm{y}_i),
    \end{split}
\end{equation}
that is a mixture of the posterior distributions of the latent order conditionally on all observations. A proposal of the form in Equation~\eqref{eq:proposal} assigns more mass to orders which are representative at least for a single observation, while having a flat proposal over the orders' space results in proposing rarely candidates which are suitable latent orders of the data. See \citet{corradin2024} for further details. Finally, at the end of each iteration, a step called \textit{acceleration step} updates all the latent orders in $\mathcal{R}^*$. Each element in $\mathcal{R}^*$ is updated with a procedure similar to the one introduced in Section~\ref{subsec:detect_cp}, but conditionally on all observations to which the latent order is assigned. Algorithm~\ref{algo:clust_cp} reported in Section~\ref{app:algorithms} of the Appendix provides the pseudocode of the clustering procedure for time series with common change points. 

\subsection{Extension to epidemiological Data} \label{subsec:epi_data}

Methods for change point detection and clustering of time dependent data with common change points in \pkg{BayesChange} are originally designed for real valued time series. However, they can be easily extended to other type of data by defining a proper likelihood function. For example, we considered the case where a generic observation $\bm{y}_i$ is a sequence of daily new infected individuals for the $i$-th population. The posterior inference procedure is similar to the one used for time series in both the change point detection and clustering methods. The only difference is that the likelihood must be defined according to the new kind of data. In \cite{corradin2024} the likelihood is derived from the discretization of a standard compartmental SIR model. The dynamic over time is described by the following differential equation system
\[
    \odv{}{t} S = - \beta(t) SI, \qquad \odv{}{t} I = \beta(t) SI - \xi(t) I,  \qquad \odv{}{t} R = \xi(t) I
\]
with the initial condition $S(0) = 0$, $I(0) = I_0$ and $R(0) = 0$. Here, $\beta(t)$ is the infection rate and $S(t)$ and $I(t)$ are respectively the number of susceptibles and infected individuals at time $t$. Hence, conditioning on a the final observational time, the density function associated to the survival function of the previous model has the following form
\begin{equation*}
        f_T(t) = - \left( \frac{1}{1-S(T)} \right) \frac{\mathrm{d}}{\mathrm{d}t}S(t) = \frac{\beta(t)S(t)I(t)}{1-S(t)}, \qquad t = 1, \dots, T.
\end{equation*}
See \cite{khudabukhsh2022projecting} and \cite{Rempaa2023Handbook} for further details on such a modelling strategy. This model is indexed by three time-dependent parameters: the vector of infection rates for a generic observation $i$, $\bm{\beta}_i = \{ \beta_{i1}, \dots, \beta_{iT} \}$, that is time dependent and observation-specific, the starting proportion of infected individuals $I_0$ that is only observation-specific and the recovery rate $\xi$ that is assumed constant over time. Since $f_T(t)$ is an intractable likelihood, these parameters cannot be integrated out analytically from the likelihood like in the time series application. Following \cite{corradin2024}, $\bm{\beta}_i$ is integrated with a Monte-Carlo procedure with the unique values sampled from a gamma distribution, $I_0$ is updated with a Metropolis-Hastings step and $\xi$ is fixed.

 \section{Package structure} \label{sec:pkg_structure}

 \pkg{BayesChange} provides two main \proglang{R} functions that call the \proglang{C++} methods, with the techniques described in Section~\ref{sec:models}. These two functions are \fct{detect\_cp} and \fct{clust\_cp} that detect change points on time series or survival functions and cluster time series or survival functions with common change points, respectively. The previous functions are designed to be intuitive to use and require only a few mandatory arguments, without the need for an in-depth understanding of the underlying models. At the same time, users with a deep knowledge of the models can specify all the parameters of the models, tailoring function calls on specific scientific interests. To this aim, the functions have been written in an object-oriented framework with two \proglang{S3} classes and methods that summarize and illustrate the final results. 

 We present in Section~\ref{sec:implementation_details} the \proglang{C++} implementation of the models, in Section~\ref{sec:user_interface} the \proglang{R} user interface and finally in Section~\ref{sec:additional_methods} the general methods that can be applied to both \proglang{S3} classes.
 
 \subsection{Implementation details} \label{sec:implementation_details}

 All functions performing posterior sampling contained in the \pkg{BayesChange} package are written in \proglang{C++}, mainly resorting to
 \pkg{Rcpp}, \pkg{RcppArmadillo} and \pkg{RcppGSL} libraries \citep{rcpp, rcpparmadillo, RcppGSL}. The latter is used mainly to sample efficiently from distributions, while the others are used to have computationally efficient samplers.

 The detection of change points is performed by three functions, \fct{detect\_cp\_uni} for univariate time series, \fct{detect\_cp\_multi} for multivariate time series and \fct{detect\_cp\_epi} for epidemiological diffusions. Both \fct{detect\_cp\_uni} and \fct{detect\_cp\_multi} consist of a \fct{for} loop that is repeated for an arbitrary number of iterations. Within each iteration, the algorithm first randomly performs a split or merge of the latent order of the data, and then, if the number of blocks is larger than one, it also performs a shuffle step. To facilitate possible future extensions of the package, most fundamental operations within a single MCMC loop are coded in independent \proglang{C++} functions. The functions \fct{AlphaSplit\_UniTS} and \fct{AlphaSplit\_MultiTS} compute the acceptance ratio of a split proposal, for univariate time series and multivariate time series, respectively. Similarly, \fct{AlphaMerge\_UniTS} and \fct{AlphaMerge\_MultiTS} compute the acceptance ratio for the proposal of a merge step, while \fct{AlphaShuffle\_UniTS} and \fct{AlphaShuffle\_MultiTS} for the proposal of a shuffle step. These functions return a value between $0$ and $1$. If the returned value is larger than a uniformly distributed random number taking values in $(0,1)$, then the proposal is accepted as a new state of latent order. Otherwise, the previous configuration of the latent order is kept as the current state. Then, at the end of each iteration, functions \fct{UpdateGamma}, \fct{UpdatePhi} and \fct{UpdateDelta} update the main parameters of the model. The function \fct{detect\_cp\_epi} has been designed in a slightly different way to optimize computational time, since computing the likelihood of survival functions arising from SIR models is more demanding. The function is composed of a \fct{for} loop. At each iteration, the function \fct{update\_I0} updates the proportion of infected individuals at time zero, assumed to be unknown, and the function \fct{update\_single\_order} updates the change points of the epidemiological diffusion.

 The \pkg{BayesChange} package performs clustering of time-dependent data with common change points with three main functions. Similarly to change points detection, \fct{clust\_cp\_uni} handles univariate time series, \fct{clust\_cp\_multi} is built for multivariate time series and \fct{clust\_cp\_epi} for epidemics. All these methods start with a function that computes an approximation of the normalization constant for the mixture in Equation~\ref{eq:proposal} with the given data. Specifically, \fct{norm\_constant\_uni}, \fct{norm\_constant\_multi} and \fct{norm\_constant\_epi} compute the normalization constant for univariate time series, multivariate time series and SIR survival functions, respectively. Then, the main \fct{for} loop run for an arbitrary number of iterations. For functions \fct{clust\_cp\_uni} and \fct{clust\_cp\_multi}, the proposals at each step are evaluated through functions \fct{AlphaSplit\_Clust} and \fct{AlphaMerge\_Clust}, analogously to the change point detection case. Before evaluating the acceptance ratio, a new latent order is assigned to each new proposed group. This is done with \fct{SplitMergeUniTS} and \fct{SplitMergeMultiTS}. These functions have the same structure of \fct{detect\_cp\_uni} and \fct{detect\_cp\_multi}, but of void type, to save memory usage. At the end of each MCMC iteration, \fct{SplitMergeUniTS} performs an acceleration step for univariate time series and \fct{SplitMergeMultiTS} for multivariate time series. The function \fct{clust\_cp\_epi} has been designed differently, for the same reason as \fct{detect\_cp\_epi}. The MCMC loop includes \fct{update\_I0}, \fct{update\_partition}, which updates the partition of the data, and \fct{update\_single\_order}, where the latter is applied to all the unique values of the latent order to perform the acceleration step. 

 \subsection{User interface} \label{sec:user_interface}

 We provide here a description of the \proglang{R} user interface of \pkg{BayesChange}. Two wrappers, \fct{detect\_cp} and \fct{clust\_cp}, have been designed to interact with the \proglang{C++} functions presented in Section~\ref{sec:implementation_details}. 

\subsubsection{Detect change points}

\begin{figure}[!h]
\centering
\begin{tikzpicture}
    \node (B) at (3, 2) [draw, rectangle] {\fct{detect\_cp}};
    \node (B1) at (-1, 0) [draw, rectangle] {\fct{detect\_cp\_uni}};
    \node (B2) at (3, 0) [draw, rectangle] {\fct{detect\_cp\_multi}};
    \node (B3) at (7, 0) [draw, rectangle] {\fct{detect\_cp\_epi}};

    \draw (B) -- (B1);
    \draw (B) -- (B2);
    \draw (B) -- (B3);

    \draw[dashed] (-4,1) -- (9,1);

    \node at (-3.5,2) {R};
    \node at (-3.5,0) {C++};

    \node at (0.75, -1.25) {\code{kernel = "ts"}};  
    \node at (7, -1.25) {\code{kernel = "epi"}};   

    \draw[decorate,decoration={brace,amplitude=10pt,mirror}] (-2, -0.5) -- (4, -0.5);
    \draw[decorate,decoration={brace,amplitude=10pt,mirror}] (6, -0.5) -- (8, -0.5);

\end{tikzpicture}
\caption{Diagram representation of function \fct{detect\_cp} call for different type of data.}\label{diag_1}
\end{figure}
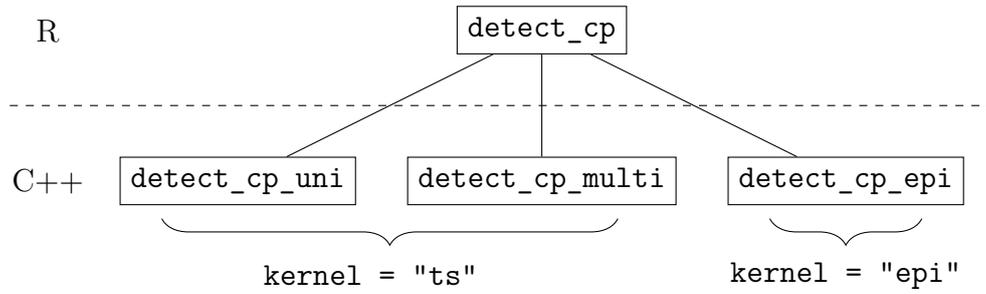

With function \fct{detect\_cp} the user can perform change point detection on univariate or multivariate time series and on epidemiological diffusions. The first argument of this function is \code{data}. It can be either a vector, if we analyze an univariate time series or an epidemiological diffusion, or a matrix, when we deal with a multivariate time series. The user specifies with the \code{kernel} argument the type of data considered in the analysis. If  \code{kernel = "ts"} the algorithm automatically detects if \code{data} is a vector or a matrix and calls the \proglang{C++} functions \fct{detect\_cp\_uni} or \fct{detect\_cp\_multi}. If \code{kernel = "epi"} then the algorithm calls \fct{detect\_cp\_epi}. A representation of this process is given by the diagram in Figure~\ref{diag_1}.

Each call of the \fct{detect\_cp} function can be tuned trough the following arguments:
\begin{itemize}
    \item \code{n_iterations}: number of MCMC iterations. 
    \item \code{n_burnin}: the number of burn-in iterations. By default \code{n_burnin = 0}. 
    \item \code{q}: the probability of performing a split at each step. By default \code{q = 0.5}. 
    \item \code{params}: a list with the parameters specific for the chosen kernel. On Table~\ref{tab:detect_cp_parameters} are detailed the arguments for both time series (univariate or multivariate setting) and epidemiological diffusions. 
    \item \code{kernel}: set \code{kernel = "ts"} if data are time series, set \code{kernel = "epi"} if data are infections from an epidemic.
    \item \code{print\_progress}: if \code{TRUE} print the progress of the algorithm. 
    \item \code{user\_seed}: the seed for the random distribution generation. 
\end{itemize}

Function \fct{detect\_cp} returns an object of class \code{"DetectCpObj"} that contains the following objects: 
\begin{itemize}
    \item \code{data}, \code{n\_iterations} and \code{n\_burnin}: these objects contain respectively the data, the number of iterations and of burn-in steps specified by the user.
    \item \code{orders}: a matrix in which are stored the latent orders sampled at each iteration. Each row corresponds to an iteration and each column to a realization of the latent order. 
    \item \code{time}: computational time of the algorithm in seconds. 
    \item \code{phi_MCMC} and \code{phi_MCMC_01}: if data are time series; these two objects respectively store the vector containing the posterior samples from the Metropolis-Hastings updates of $\phi$, and a binary vector indicating whether the proposed value of $\phi$ was accepted ($1$) or rejected ($0$) at each iteration.
    \item \code{sigma_MCMC} and \code{sigma_MCMC_01}: if the data are time series; these two objects respectively store the vector containing the posterior samples from the Metropolis-Hastings updates of $\phi$, and a binary vector indicating whether the proposed value of $\sigma$ was accepted ($1$) or rejected ($0$) at each iteration.
    \item \code{delta_MCMC}: if data are time series; a vector with posterior sample of $\delta$. 
    \item \code{I0_MCMC} and \code{I0_MCMC_01}: if data are time series; these two objects respectively store the vector containing the posterior samples from the Metropolis-Hastings updates of $I_0$, and a binary vector indicating whether the proposed value of $\sigma$ was accepted ($1$) or rejected ($0$) at each iteration.
    \item \code{kernel\_ts} and \code{kernel\_epi}: boolean objects equal to \code{TRUE} if data are respectively time series or infections of an epidemic diffusion.
    \item \code{univariate\_ts}: if data are time series; a boolean object equal to \code{TRUE} if data is an univariate time series, \code{FALSE} if it is a multivariate time series. 
\end{itemize}

\begin{table}[h]
    \centering
    \renewcommand{\arraystretch}{1.4} 
    \begin{tabular}{c l l}
        \hline
        \textbf{Model} & \textbf{Argument} & \textbf{Interpretation} \\
        \hline
        \multirow{6}{*}{\shortstack[c]{Univariate \\ Time Series}} 
        & \code{a} & \multirow{3}{*}{parameters of the prior $\text{N}(0,(c\lambda)^{-1})\text{Ga}(a,b)$ for $\mu$ and $\lambda$} \\
        & \code{b} &  \\
        & \code{c} &  \\
        & \code{prior_var_phi} & variance $\sigma^2_\phi$ in the proposal $N(0,\sigma^2_\phi)$ for the estimate of $\phi$ \\
        & \code{prior_delta_c} & \multirow{2}{*}{parameters of the full conditional distribution of $\delta$}\\
        & \code{prior_delta_d} & \\
        \hline
        \multirow{7}{*}{\shortstack[c]{Multivariate \\ Time Series}} & \code{m_0} & \multirow{4}{*}{parameters of the prior $NIW(\bm{m}_0, k_0, \nu_0,S_0)$ for $\bm{\mu}$ and $\Lambda$}\\
        & \code{k_0} &  \\
        & \code{nu_0} &  \\
        & \code{S_0} & \\
        & \code{prior_var_phi} & variance $\sigma^2_\phi$ in the proposal $N(0,\sigma^2_\phi)$ for the estimate of $\phi$\\
        & \code{prior_delta_c} & \multirow{2}{*}{parameters of the full conditional distribution of $\delta$}\\
        & \code{prior_delta_d} & \\
        \hline
        \multirow{5}{*}{\shortstack[c]{Epidemic \\ Diffusions}} & \code{M} & number of Monte Carlo iterations for the likelihood integration \\
        & \code{xi} & recovery rate $\xi$  \\
        & \code{a0} & \multirow{2}{*}{\shortstack[l]{parameters of the Gamma proposal used to integrate out \\ the infection rates}} \\
        & \code{b0} &  \\
        & \code{I0_var} & variance in the normal $N(0,\sigma^2_{I_0})$ proposal for updating $I_0$ \\
        \hline
    \end{tabular}
    \caption{Parameters for the list of arguments in \code{params} of \fct{detect\_cp}}
    \label{tab:detect_cp_parameters}
\end{table}

\begin{table}[h]
    \centering
    \renewcommand{\arraystretch}{1.4} 
    \begin{tabular}{c l l}
        \hline
        \textbf{Model} & \textbf{Argument} & \textbf{Interpretation} \\
        \hline
        \multirow{4}{*}{\shortstack[c]{Univariate \\ Time Series}}
        & \code{a} & \multirow{3}{*}{parameters of the prior $\text{N}(0,(c\lambda)^{-1})\text{Ga}(a,b)$ for $\mu$ and $\lambda$} \\
        & \code{b} &  \\
        & \code{c} &  \\
        & \code{phi} & correlation parameter in $OU(\mu,\lambda,\phi)$ \\
        \hline
        \multirow{5}{*}{\shortstack[c]{Multivariate \\ Time Series}} & \code{k_0} & \multirow{4}{*}{parameters of the prior $NIW(\bm{m}_0, k_0, \nu_0,S_0)$ for $\bm{\mu}$ and $\Lambda$}\\
        & \code{nu_0} &  \\
        & \code{S_0} &  \\
        & \code{m_0} &  \\
        & \code{phi} & correlation parameter in $OU(\bm{\mu},\lambda,\phi)$ \\
        \hline
        \multirow{6}{*}{\shortstack[c]{Epidemic \\ Diffusions}} & \code{M} & number of Monte Carlo iterations for the likelihood integration \\
        & \code{xi} & recovery rate $\xi$  \\
        & \code{a0} & \multirow{2}{*}{\shortstack[l]{parameters of the Gamma proposal used to integrate out \\ the infection rates}} \\
        & \code{b0} &  \\
        & \code{I0_var} & variance in the normal $N(0,\sigma^2_{I_0})$ proposal for updating $I_0$ \\
        & \code{avg_blk} & average number of blocks when random orders are generated \\
        \hline
    \end{tabular}
    \caption{Parameters for the list of arguments \code{params} of \fct{clust\_cp}}
    \label{tab:clust_cp_parameters}
\end{table}

\subsubsection{Clustering data with common change points}

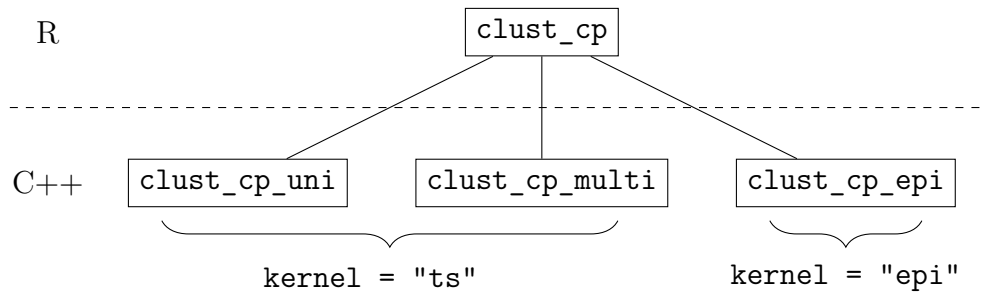
\begin{figure}
\centering
\begin{tikzpicture}
    \node (B) at (3, 2) [draw, rectangle] {\fct{clust\_cp}};
    \node (B1) at (-1, 0) [draw, rectangle] {\fct{clust\_cp\_uni}};
    \node (B2) at (3, 0) [draw, rectangle] {\fct{clust\_cp\_multi}};
    \node (B3) at (7, 0) [draw, rectangle] {\fct{clust\_cp\_epi}};

    \draw (B) -- (B1);
    \draw (B) -- (B2);
    \draw (B) -- (B3);

    \draw[dashed] (-4,1) -- (9,1);

    \node at (-3.5,2) {R};
    \node at (-3.5,0) {C++};

    \node at (0.75, -1.25) {\code{kernel = "ts"}};  
    \node at (7, -1.25) {\code{kernel = "epi"}};   

    \draw[decorate,decoration={brace,amplitude=10pt,mirror}] (-2, -0.5) -- (4, -0.5);
    \draw[decorate,decoration={brace,amplitude=10pt,mirror}] (6, -0.5) -- (8, -0.5);

\end{tikzpicture}

\caption{Diagram representation of function \fct{clust\_cp} call for different type of data.}\label{diag_2}
\end{figure}

\fct{clust\_cp} includes the three methods to perform clustering of time-dependent observations with common change points. The structure is similar to that of \fct{detect\_cp}. The function still includes arguments \code{data}, \code{n\_iterations}, \code{n\_burnin}, \code{q}, \code{kernel}, \code{print\_progress} and \code{user_seed}, with slight differences for \code{data} and \code{q}. For univariate time series or epidemic diffusions, the \code{data} argument is a matrix where each row is an observation and each column a realization. If data are multivariate time series, \code{data} is a multidimensional array, where each slice is a matrix with rows denoting the dimensions and columns the realizations. The parameter \code{q} also denotes the probability of a split, but here for the acceleration step and in the split-merge proposal step of the algorithm. The argument \code{params} contains parameters specific for each type of data, detailed in Table~\ref{tab:clust_cp_parameters}. Based on the specified \code{kernel} and the data type, \fct{clust\_cp} calls \fct{clust\_cp\_uni}, \fct{clust\_cp\_multi} or \fct{clust\_cp\_epi}, according to the diagram in Figure~\ref{diag_2}. 
Further, the new arguments in \fct{clust\_cp} are the following:
\begin{itemize}
    \item \code{alpha_SM}: the parameter value of the symmetric Dirichlet distribution in Equation~\ref{mod:spec}. 
    \item \code{B}: the number of latent orders randomly generated when the approximation of the normalizations constant is computed. 
    \item \code{L}: the number of split-and-merge steps performed to propose a latent order for the new groups. 
\end{itemize}

The output of \fct{clust\_cp} is an object of class \code{"ClustCPObj"}. Besides the inputs of the user, saved in \code{data}, \code{n_iterations} and \code{n_burnin} , it contains the following elements:
\begin{itemize}
    \item \code{clust}: a matrix where each row corresponds to the latent partition of the data, returned at each iteration of the MCMC algorithm. 
    \item \code{orders}: an array where each element is a matrix with the unique latent orders of the proposed partition, specific for each iteration. 
    \item \code{time}: computational time of the algorithm in seconds. 
    \item \code{norm_vec}: the vector with the observation-specific contribution of the approximated normalization constant. 
    \item \code{I0_MCMC} and \code{I0_MCMC_01}: if the data are epidemic diffusions, these two objects respectively store the vector containing the posterior samples from the Metropolis-Hastings updates of $I_0$, and a binary vector indicating whether the proposed value of $\sigma$ was accepted ($1$) or rejected ($0$) at each iteration.
    \item \code{kernel\_ts} and \code{kernel\_epi}: boolean objects equal to \code{TRUE} if data are respectively time series or epidemic diffusions.
    \item \code{univariate\_ts}: only if \code{kernel\_ts = TRUE}, if \code{TRUE} time series are univariate, \code{FALSE} if they are multivariate. 
\end{itemize}

 \subsection{Generic methods and additional functions} \label{sec:additional_methods}

 \pkg{BayesChange} provides \proglang{S3} methods for both \code{"DetectCpObj"} and \code{"ClustCpObj"} objects. These methods summarize the output of functions \code{detect\_cp} and \code{clust\_cp}, give information about the algorithm, provide a point estimation method and a graphical illustration of the results. 

 The first method is \fct{print}, it returns a message that says which kind of data have been analyzed and what type of algorithm has been run. It says if the algorithm is for change point detection or clustering of time dependent data, and also if data are univariate time series, multivariate time series or epidemic diffusions. Method \fct{summary} returns the same information of \fct{print} and in addition the number of iterations, of burn-in steps and the computational time in seconds. 
 Method \fct{posterior\_estimate} provides a point estimate by making use of package \pkg{salso}. This package is based on the search algorithm \code{salso} by \cite{dahl2022search} which provides a point estimate for a random partition. We included this dependence because the \pkg{salso} package implements several loss functions and embeds other popular methods for estimating latent partitions as special cases. The first argument of \fct{posterior\_estimate} is an object of class \code{"DetectCpObj"} or \code{"ClustCpObj"}, the other arguments are the same of function \code{salso}. When performing change point detection, i.e. for \code{"DetectCpObj"} objects, the function returns an estimate of their locations. When clustering time-dependent quantites, i.e. for \code{"ClustCpObj"} objects, the function returns an estimate of the latent partition of the data. Finally, the \fct{plot} method has been extended resorting to the \pkg{ggplot2} package. It takes as arguments the same of \fct{posterior\_estimate}, since it first computes a point estimate before providing a graphical representation. 

 \pkg{BayesChange} includes also the function \fct{sim\_epi\_data}. Such a function generates synthetic survival data using the Doob–Gillespie algorithm. See \cite{anderson2015stochastic} for further details. The output of the function is a vector with the simulated infection times. The arguments of \fct{sim\_epi\_data} are the following:
 \begin{itemize}
     \item \code{S0}: number of individuals in the population.
     \item \code{I0}: number of infected individuals at time zero. 
     \item \code{max\_time}: maximum observed time. 
     \item \code{beta\_vec}: vector of time-dependent infection rates.
     \item \code{xi_0}: recovery rate.
     \item \code{user_seed}: the seed for the random distribution generation.
 \end{itemize}

\section{Illustrations} \label{sec:illustrations}


In this section, we guide the reader through the use of the main methodologies provided by the \pkg{BayesChange} package. We show how to detect change points in time series and epidemic diffusions with \fct{detect\_cp}, and how to cluster time-dependent data that share the same change points with \fct{clust\_cp}. To illustrate all functions in a reasonable computational time, we decide here to implement \pkg{BayesChange} on synthetic data. For each function, we provide the code to generate synthetic data, which helps to illustrate which kind of input object is required to run each specific algorithm. 
In the context of epidemic diffusions, we also show how to simulate data with \fct{sim\_epi\_data}. For each illustration, we provide the code and the plot of the final estimate.

\subsection{Change points detection on time series and survival functions}
\begin{figure}[!h]
    \centering
    \includegraphics[width=1\linewidth]{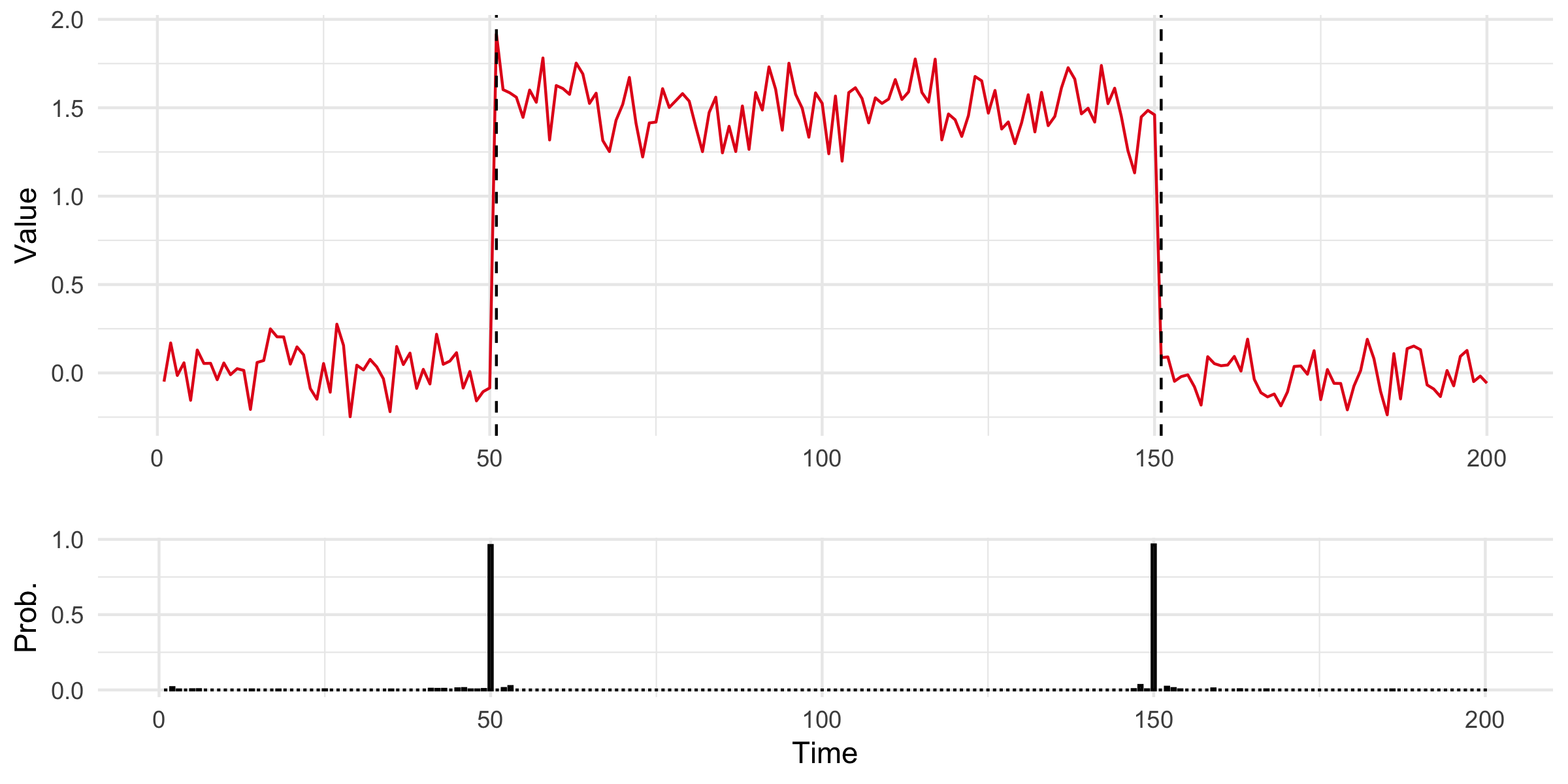}
    \caption{Detected change points on univariate synthetic time series. The dashed lines represent the estimated change points.}
    \label{fig:uni_detect_cp}
\end{figure}

\begin{figure}[!h]
    \centering
    \includegraphics[width=1\linewidth]{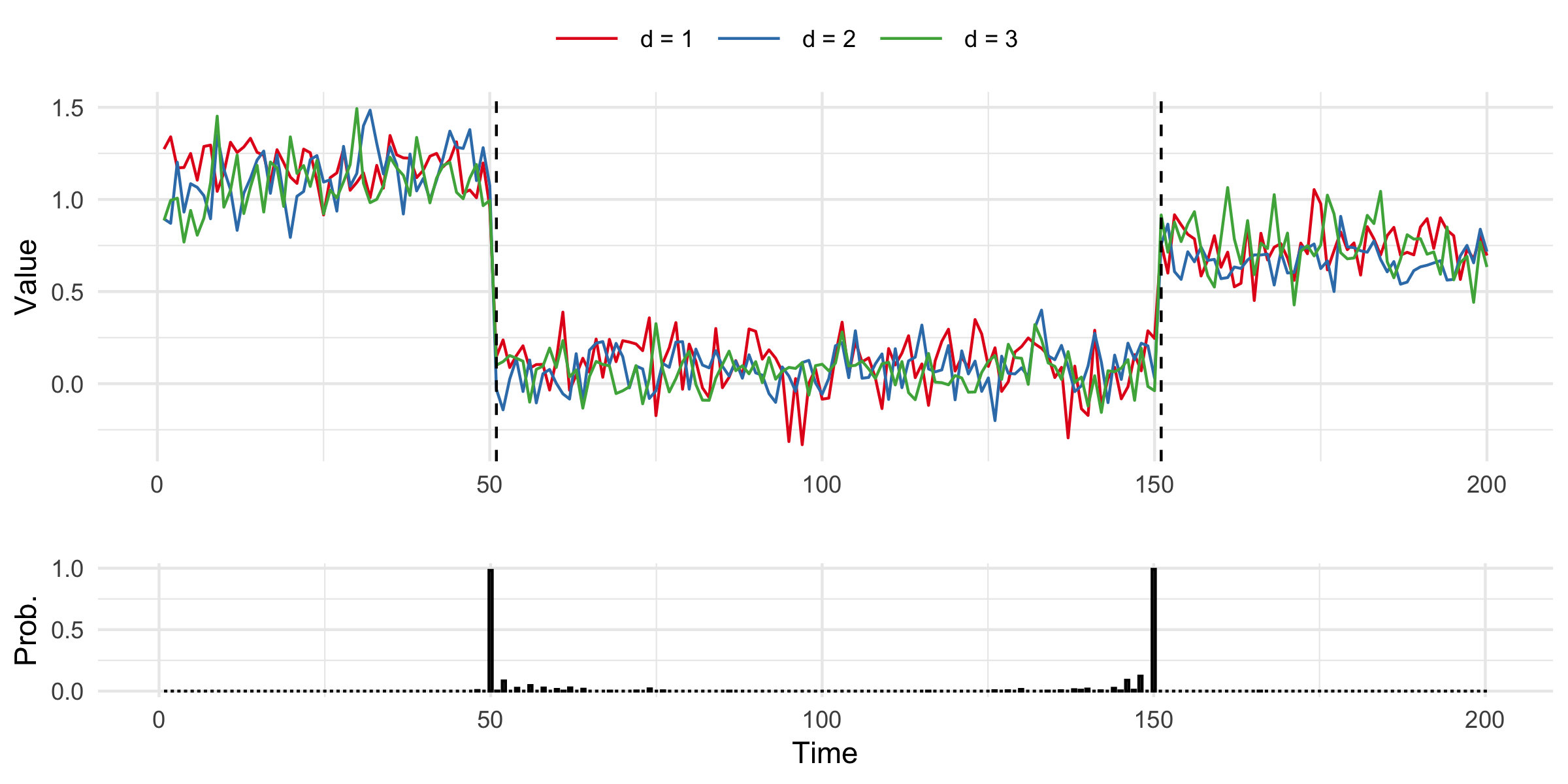}
    \caption{Detected change points on multivariate synthtetic time series. The dashed lines represent the position of the estimated change points. Each color denotes a different dimension of the time series.}
    \label{fig:multi_detect_cp}
\end{figure}

At first, we show how to run an analysis with the \fct{detect\_cp} function on all different type of data \pkg{BayesChange} can handle. Since the models implemented in \pkg{BayesChange} are capable of detecting changes in both the mean and the variance, we generate synthetic data from an autoregressive Gaussian process in which both the local mean and variance change at each change point. We simulate a time series with $200$ realizations and two change points, respectively located at times $51$ and $151$. 
\begin{CodeChunk}
\begin{CodeInput}
R> data <- as.numeric()
R> data[1] <- rnorm(1, mean = 0, sd = 0.13)
R> for(i in 2:50){
R>  data[i] <- 0.1 * data[i-1] + (1 - 0.1) * 0 + 
R>      rnorm(1, mean = 0, sd = (1 - 0.1^2) * 0.13)}
R> data[51] <- rnorm(1, mean = 1.5, sd = 0.15)
R> for(i in 52:150){
R>  data[i] <- 0.1 * data[i-1] + (1 - 0.1) * 1.5 + 
R>      rnorm(1, mean = 0, sd = (1 - 0.1^2) * 0.15)}
R> data[151] <- rnorm(1, mean = 0, sd = 0.12)
R> for(i in 152:200){
R>  data[i] <- 0.1 * data[i-1] + (1 - 0.1) * 0 + 
R>      rnorm(1, mean = 0, sd = (1 - 0.1^2) * 0.12)
R> }
\end{CodeInput}
\end{CodeChunk}
Before running the algorithm, we need to set the specific parameter for the univariate model by defining a list with the arguments of the Normal-Gamma prior, as shown in Section \ref{sec:user_interface}. 
\begin{CodeChunk}
\begin{CodeInput}
R> params_uni <- list(a = 1, b = 1, c = 1, prior_var_phi = 0.1, 
+                  prior_delta_c = 1, prior_delta_d = 1)
\end{CodeInput}
\end{CodeChunk}
The algorithm is run by calling function \fct{detect\_cp}. We include the list \code{params\_uni} in the argument \code{params} and we also specify other general arguments, common to both the time series and epidemiological frameworks. Specifically, we specify the number of iterations, the number of burn-in steps and the probability of performing a split at each iteration. 
\begin{CodeChunk}
\begin{CodeInput}
R> out <- detect_cp(data, n_iterations = 10000, n_burnin = 5000, q = 0.25, 
+           params = params_uni, kernel = "ts")
R> print(out)
\end{CodeInput}
\begin{CodeOutput}
DetectCpObj object
Type: change points detection on univariate time series
\end{CodeOutput}
\end{CodeChunk}
To get a posterior estimate of the change points we use \fct{posterior\_estimate}, we leave all arguments to default values and set \textit{Binder} as loss function. 
\begin{CodeChunk}
\begin{CodeInput}
R> cp_est <- posterior_estimate(out, loss = "binder")
\end{CodeInput}
\end{CodeChunk}
The output of \code{posterior\_estimate} is a sequence of number that represents the allocation of each realization to a block. In order to get the position of the change points it is sufficient to print the cumulative sum of the frequency table of the vector, remove the last element and sum one. 
\begin{CodeChunk}
\begin{CodeInput}
R> cumsum(table(cp_est))[-length(table(cp_est)] + 1
\end{CodeInput}
\begin{CodeOutput}
  1   2  
 51 151
\end{CodeOutput}
\end{CodeChunk}
Finally we graphically represent the detected change points along with the time series with \fct{plot}. This method also provides, by setting \code{plot\_freq = TRUE}, the histogram with the frequency of times that each realization has been detected as change point in the MCMC chain. In Figure~\ref{fig:uni_detect_cp} is reported the output generated by the following code
\begin{CodeChunk}
\begin{CodeInput}
R> plot(x = out, loss = "binder", plot_freq = TRUE)
\end{CodeInput}
\end{CodeChunk}
In the multivariate scenario, we need to define a matrix in which each row corresponds to a dimension of the time series. We consider the same number of realizations and the same change point locations as in the univariate example. A synthetic multivariate time series is generated with the following code:
\begin{CodeChunk}
\begin{CodeInput}
R> data <- matrix(NA, nrow = 3, ncol = 200)
R> data[1, 1] <- rnorm(1, mean = 1.20, sd = 0.12)
R> data[2, 1] <- rnorm(1, mean = 1.15, sd = 0.15)
R> data[3, 1] <- rnorm(1, mean = 1.10, sd = 0.14)
R> for(i in 2:50){
R>  data[1, i] <- 0.1 * data[1, i-1] + (1 - 0.1) * 1.20 + 
R>      rnorm(1, mean = 0, sd = (1 - 0.1^2) * 0.12)
R>  data[2, i] <- 0.1 * data[2, i-1] + (1 - 0.1) * 1.15 + 
R>      rnorm(1, mean = 0, sd = (1 - 0.1^2) * 0.15)
R>  data[3, i] <- 0.1 * data[3, i-1] + (1 - 0.1) * 1.10 + 
R>      rnorm(1, mean = 0, sd = (1 - 0.1^2) * 0.14)
R> }
R> data[1, 51] <- rnorm(1, mean = 0.06, sd = 0.14)
R> data[2, 51] <- rnorm(1, mean = 0.07, sd = 0.12)
R> data[3, 51] <- rnorm(1 ,mean = 0.08, sd = 0.10)
R> for(i in 52:150){
R>  data[1, i] <- 0.1 * data[1, i-1] + (1 - 0.1) * 0.06 + 
R>      rnorm(1, mean = 0, sd = (1 - 0.1^2) * 0.14)
R>  data[2, i] <- 0.1 * data[2, i-1] + (1 - 0.1) * 0.07 + 
R>      rnorm(1, mean = 0, sd = (1 - 0.1^2) * 0.12)
R>  data[3, i] <- 0.1 * data[3, i-1] + (1 - 0.1) * 0.08 + 
R>      rnorm(1, mean = 0, sd = (1 - 0.1^2) * 0.10)
R> }
R> data[1, 151] <- rnorm(1, mean = 0.72, sd = 0.13)
R> data[2, 151] <- rnorm(1, mean = 0.69, sd = 0.10)
R> data[3, 151] <- rnorm(1, mean = 0.75, sd = 0.14)
R> for(i in 152:200){
R>  data[1, i] <- 0.1 * data[1, i-1] + (1 - 0.1) * 0.72 + 
R>      rnorm(1, mean = 0, sd = (1 - 0.1^2) * 0.13)
R>  data[2, i] <- 0.1 * data[2, i-1] + (1 - 0.1) * 0.69 + 
R>      rnorm(1, mean = 0, sd = (1 - 0.1^2) * 0.10)
R>  data[3, i] <- 0.1 * data[3, i-1] + (1 - 0.1) * 0.75 + 
R>      rnorm(1, mean = 0, sd = (1 - 0.1^2) * 0.14)
R> }
\end{CodeInput}
\end{CodeChunk}
Similarly to the previous case, we need to specify the parameters of the Normal-Inverse Wishart distribution.
\begin{CodeChunk}
\begin{CodeInput}
R> params_multi <- list(m_0 = rep(0, 3), k_0 = 1, nu_0 = 5, 
+                   S_0 = diag(0.1, 3, 3), prior_var_phi = 0.1, 
+                   prior_delta_c = 1, prior_delta_d = 1)
\end{CodeInput}
\end{CodeChunk}
We run the algorithm providing \code{params_multi} as argument, secifying also other parameters as in the univariate scenario.  
\begin{CodeChunk}
\begin{CodeInput}
R> out <- detect_cp(data, n_iterations = 10000, n_burnin = 5000, q = 0.5, 
+           params = params_multi, kernel = "ts")
R> print(out)
\end{CodeInput}
\begin{CodeOutput}
DetectCpObj object
Type: change points detection on multivariate time series
\end{CodeOutput}
\end{CodeChunk}
Figure \ref{fig:multi_detect_cp} shows the graphical representation of the estimated change points with the \textit{Binder loss function}, along with the observed data and the probability of having a change in the time domain. Each color corresponds to a specific dimension. 

\begin{figure}[!h]
    \centering
    \includegraphics[width=1\linewidth]{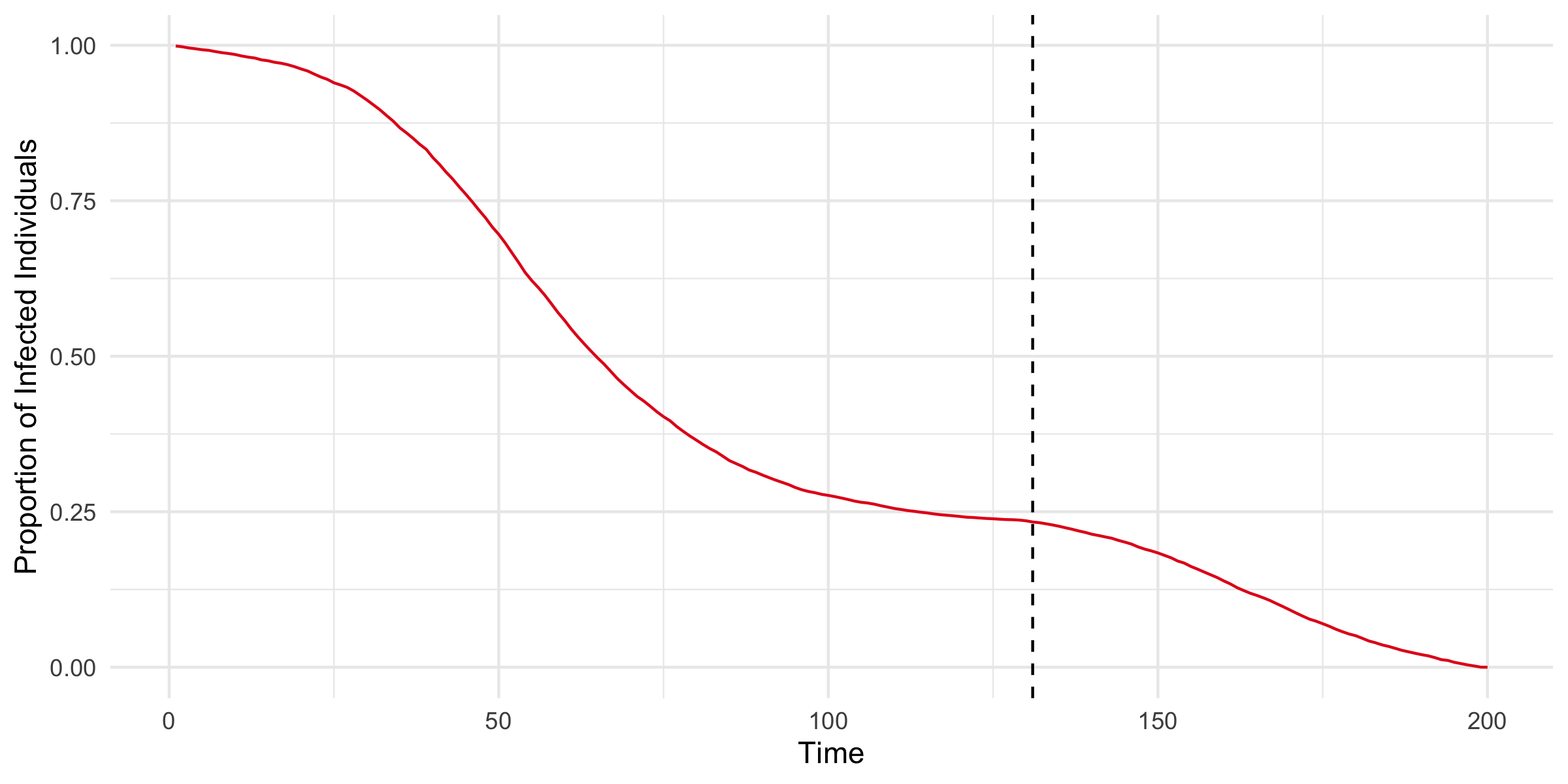}
    \caption{Detected change points on a synthetic epidemiological diffusion. The dashed line represents the position of the change point estimated by the model.}
    \label{fig:epi_detect_cp}
\end{figure}

To show how to detect change points on a survival function, we start by generating synthetic infection times on an interval of time $(0, 200)$. We assume a population of $10\,000$ individuals, of which $50$ are infected at time $0$. The recovery rate is set at $1/8$ and a change point occurs at time $131$, where the infection rate switches from $0.2$ to $0.55$. With the following code we create the vector of infection rates 
\begin{CodeChunk}
\begin{CodeInput}
R> betas <- c(rep(0.2, 130), rep(0.55, 70))
\end{CodeInput}
\end{CodeChunk}
The function \fct{sim\_epi\_data} returns a vector of continuous infection times. Since the input of \fct{detect\_cp} requires discrete time points, we round the output of \fct{sim\_epi\_data} and compute the number of new infections at each time. The result is a one-column matrix, \code{inf\_count}, where each entry represents the number of new infections at the corresponding time indicated by the row.
\begin{CodeChunk}
\begin{CodeInput}
R> inf_times <- sim_epi_data(S0 = 10000, I0 = 50, max_time = 200, 
+                beta_vec = betas, xi_0 = 1/8)
R> inf_times <- table(floor(inf_times))
R> inf_count <- matrix(0, 1, 200)
R> inf_count[as.numeric(names(inf_times)), 1] <- inf_times
\end{CodeInput}
\end{CodeChunk}
After specifying the specific parameters of the survival function on list \code{params\_epi}, we run the algorithm.
\begin{CodeChunk}
\begin{CodeInput}
R> out <- detect_cp(data = inf_count, n_iterations = 5000, n_burnin = 2000,  
+           q = 0.25, params = params_epi, kernel = "epi")
R> print(out)
\end{CodeInput}
\begin{CodeOutput}
DetectCpObj object
Type: change points detection on an epidemic diffusion
\end{CodeOutput}
\end{CodeChunk}
Figure \ref{fig:epi_detect_cp} shows the output obtained with method \fct{plot}, the change point is shown on the empirical survival function of the data generating infection times model. 

\subsection{Cluster time dependent data with common change points} \label{sec:clust_data_app}
\begin{figure}[!h]
    \centering
    \includegraphics[width=1\linewidth]{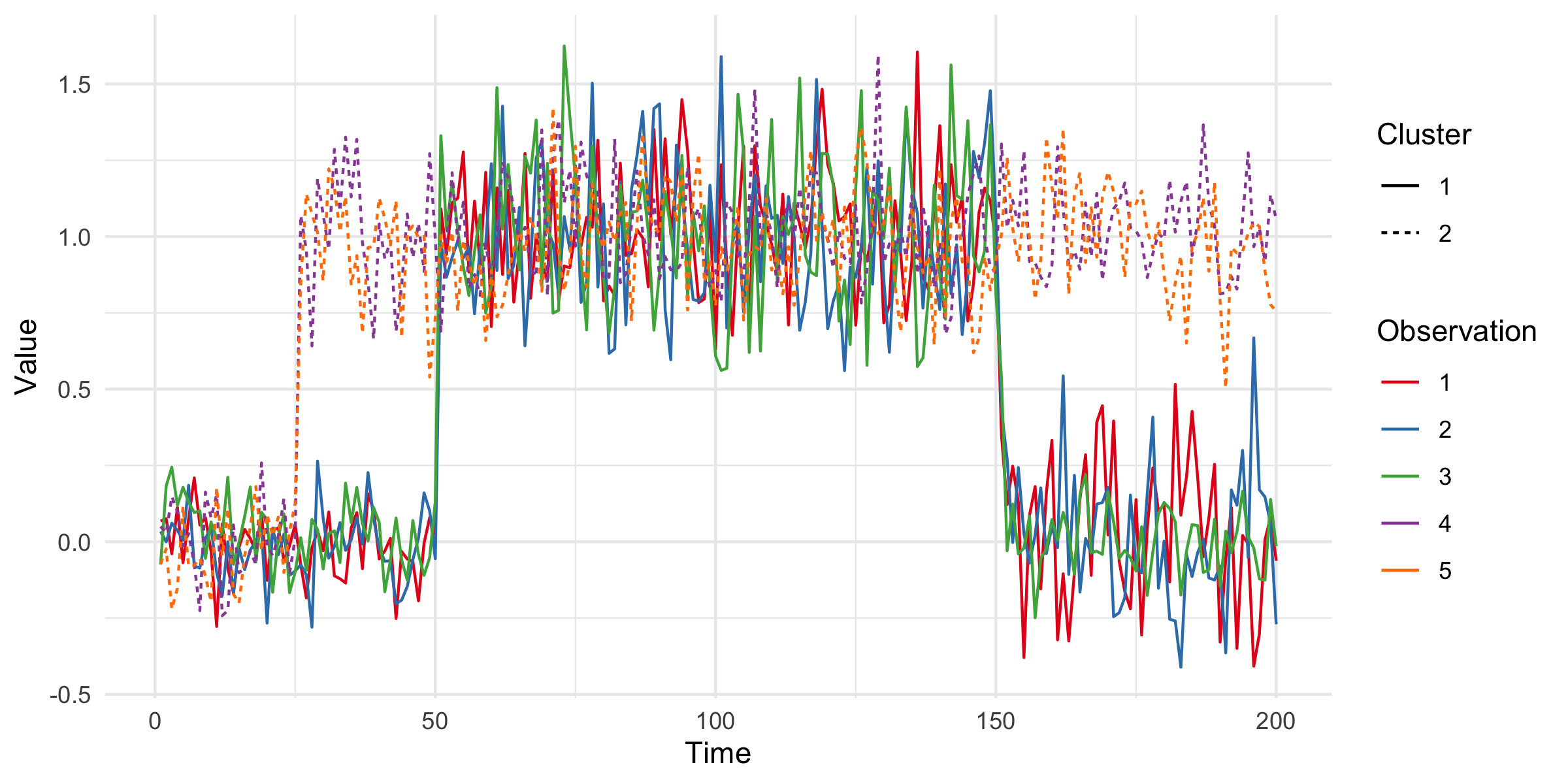}
    \caption{Clustering univariate time series with common change points. Different colors denote observations, different line types denote the cluster assignments.}
    \label{fig:uni_clust_cp}
\end{figure}

\begin{figure}[]
    \centering
    \includegraphics[width=1\linewidth]{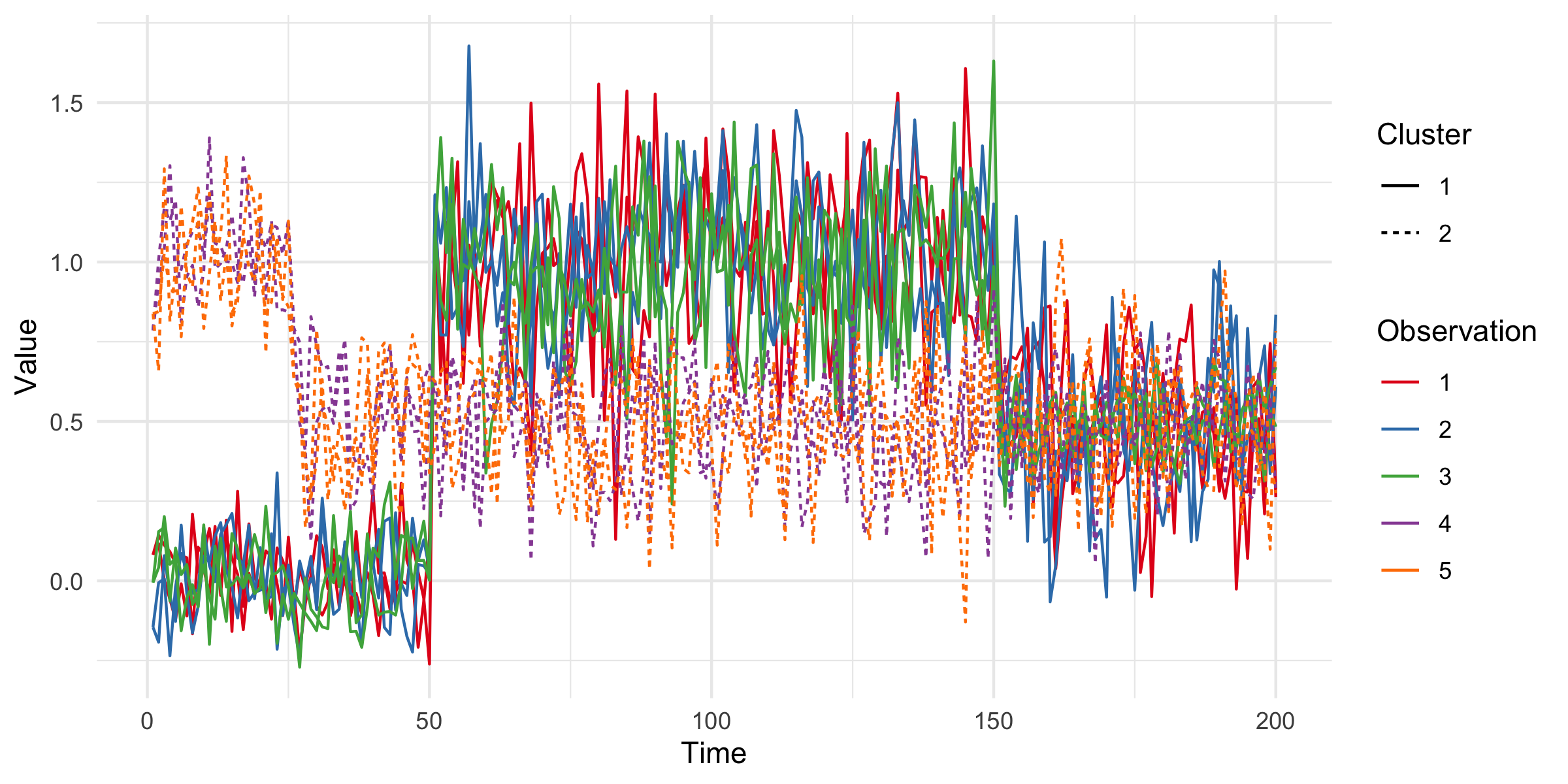}
    \caption{Clustering multivariate time series with common change points. Different color denote observations, different line types denote the cluster assignments.}
    \label{fig:multi_clust_cp}
\end{figure}

\begin{figure}[!h]
\centering
    \includegraphics[width=1\linewidth]{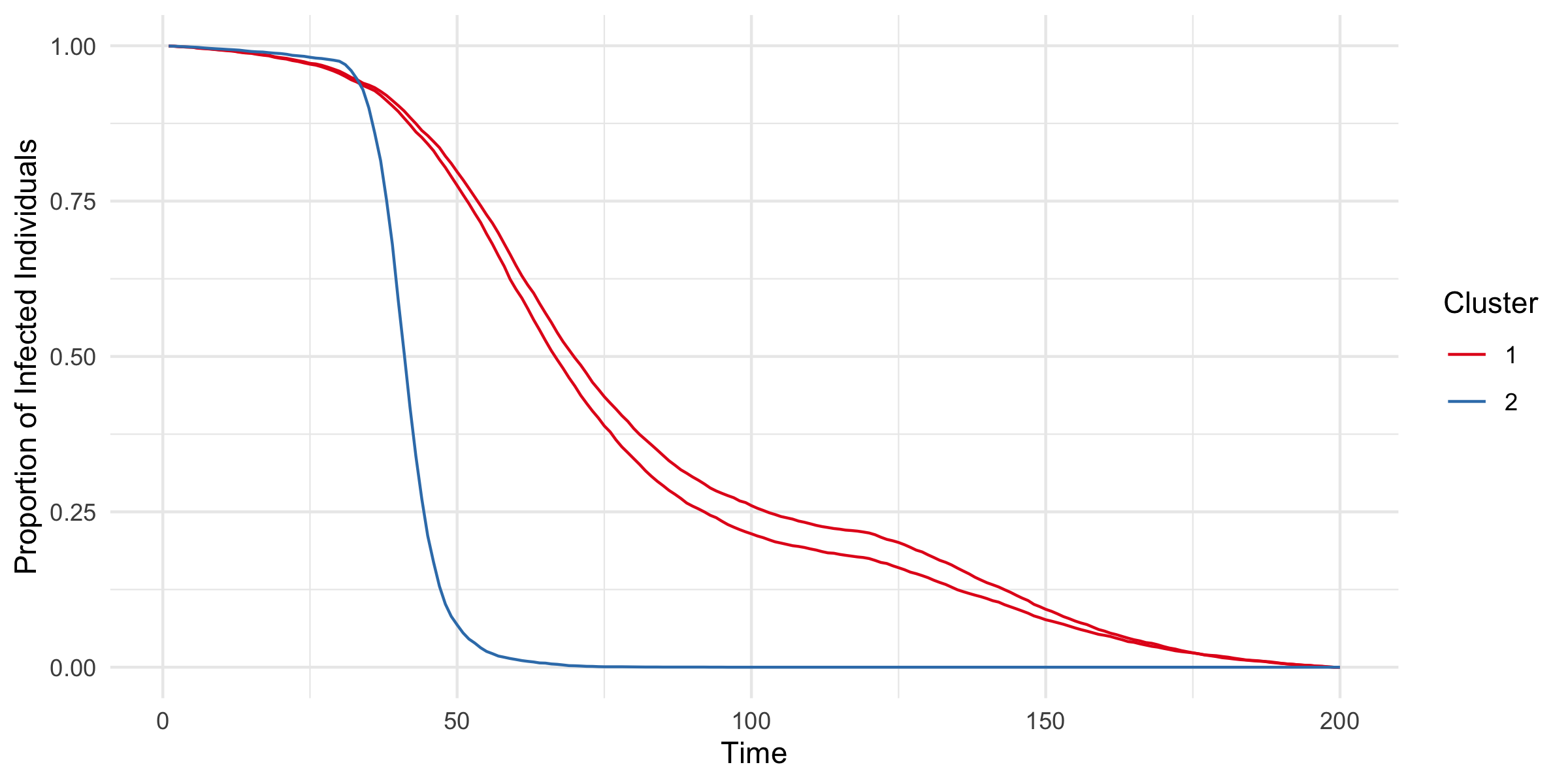}
    \caption{Clustering of survival functions with common change points. Different colors denote the cluster assignments.} 
    \label{fig:surv_clust_cp}
\end{figure}

Function \fct{clust\_cp} clusters time series or epidemic diffusions with common change points. The wrapper will perform the proper algorithm, depending on the type of input data. First, we cluster univariate time series. Data are in the form of a matrix, where each row is a time series and each column a time instant. We consider $5$ time series with $200$ realizations each, where observations are divided in two groups, of size $3$ and $2$ respectively. For the first group, we assume that data share two change points, one at time $51$ and one at time $151$. For the second group, there is only one change point at time $26$. The code for generating these data is reported in Appendix \ref{app:code}. 
To run the algorithm, it is mandatory to specify which kind of kernel describes the data, here \code{kernel = "ts"}. We specify in a list the required tuning parameters, which are the same of univariate change points detection, except for the autoregressive coefficient $\phi$, here assumed to be fixed. 
\begin{CodeChunk}
\begin{CodeInput}
R> params_uni <- list(a = 0.1, b = 1, c = 1, phi = 0.1)
\end{CodeInput}
\end{CodeChunk}
We run the algorithm with the following code: 
\begin{CodeChunk}
\begin{CodeInput}
R> out <- clust_cp(data, n_iterations = 10000, n_burnin = 5000,
+                  L = 1, q = 0.5, B = 10000, params = params_uni, 
+                  kernel = "ts")
R> print(out)
\end{CodeInput}
\begin{CodeOutput}
ClustCpObj object
Type: clustering univariate time series with common change points   
\end{CodeOutput}
\end{CodeChunk}
As shown by the \fct{print} method, the algorithm detects that time series are univariate. We estimate the latent partition of the data by calling the \fct{posterior\_estimate} function. The output is a vector of numbers denoting the clustering allocation of each observation. 
\begin{CodeChunk}
\begin{CodeInput}
R> posterior_estimate(out, loss = "binder)
R> plot(out, loss = "binder)
\end{CodeInput}
\begin{CodeOutput}
[1] 1 1 1 2 2
\end{CodeOutput}
\end{CodeChunk}
The method \fct{plot} shows a graphical representation of the estimated latent partition of the data, which is shown in Figure~\ref{fig:uni_clust_cp}. Different colors denote different observations, while different line types denote different clusters.
\begin{CodeChunk}
\begin{CodeInput}
R> plot(out, loss = "binder)
\end{CodeInput}
\end{CodeChunk}
When considering multivariate time series, each observation is a multivariate time series, we need to define data as an array. Each slice of the array, denoted as the third index, is a matrix that corresponds to an observation. For each observation, the number of rows corresponds to the dimensions of the time series, while the number of columns is the number of observational times. Here we sample $5$ time series, each one observed at $200$ times and with $2$ dimensions. The code to generate data is available in Appendix~\ref{app:code}. We create a list specifying the parameters for the multivariate kernel function. 
\begin{CodeChunk}
\begin{CodeInput}
R> params_multi <- list(m_0 = rep(0,2),  k_0 = 1, nu_0 = 5, 
+                       S_0 = diag(1, 2, 2), phi = 0.1)
\end{CodeInput}
\end{CodeChunk}
We run the algorithm calling the \fct{detect\_cp} function. Figure \ref{fig:multi_clust_cp} shows a graphical representation of the posterior point estimate of the data clustering.
\begin{CodeChunk}
\begin{CodeInput}
R> out <- clust_cp(data = data, n_iterations = 10000, n_burnin = 5000,
+                  L = 1, B = 10000, params = params_multi, kernel = "ts")
R> print(out)
\end{CodeInput}
\begin{CodeOutput}
ClustCpObj object
Type: clustering multivariate time series with common change points    
\end{CodeOutput}
\end{CodeChunk}
Finally, we show how to apply \fct{clust\_cp} to epidemic diffusions. Input data are of matrix form, where rows denote different populations and columns the observational times. Each entry is the number of new infected individuals at a specific time in a specific population. We generate infection times from $3$ populations and $50$ time instants. We assume two groups, the first with two observations and a change point at time $121$ and the second with one observation and a change point at time $31$. To generate these data, we use \fct{sim\_epi\_data}. For each population, we consider $10\,000$ individuals, of which $20$ are infected at time $0$, and a recovery rate equal to $1/8$. 
\begin{CodeChunk}
\begin{CodeInput}
R>  data <- matrix(0, nrow = 3, ncol = 200)
R>  inf_times <- list()
R>
R>  betas <- list(c(rep(0.211, 120),rep(0.55, 80)),
+                 c(rep(0.215, 120),rep(0.52, 80)),
+                 c(rep(0.193, 30),rep(0.53, 170)))
R>
R> for(i in 1:3){
R>  inf_times[[i]] <- sim_epi_data(10000, 20, 200, betas[[i]], 1/8)
R>  inf_times[[i]] <- table(floor(inf_times[[i]]))
R>  data[i,as.numeric(names(inf_times[[i]]))] <- inf_times[[i]]
R> }        
\end{CodeInput}
\end{CodeChunk}
We run the \fct{clust\_cp} function for epidemic diffusions. Specifically, we set \code{kernel = "epi"}. Similarly to before, we create a list with the specific parameters of the algorithm. Specifically, the number of Monte Carlo iterations for the likelihood integration, the recovery rate, the parameters of the weights distribution, the parameters of the Gamma and the Normal proposals, and the average number of blocks when a latent order is randomly generated.
\begin{CodeChunk}
\begin{CodeInput}
R> params_epi <- list(M = 1000, xi = 1/8, alpha_SM = 1, 
+                     a0 = 3, b0 = 10, I0_var = 0.1, avg_blk = 5)  
R> out <- clust_cp(data, n_iterations = 5000, n_burnin = 2000,
+                  L = 1, B = 1000, params = params_epi, kernel = "epi")
R> print(out)  
\end{CodeInput}
\begin{CodeOutput}
ClustCpObj object
Type: clustering epidemic diffusions with common change points
\end{CodeOutput}
\end{CodeChunk}
Figure \ref{fig:surv_clust_cp} shows the posterior point estimate of the clusters along with the empirical survival functions of the epidemic diffusions, specific for each population. 


\section{Summary and discussion} \label{sec:summary}

We presented in this paper \pkg{BayesChange}, an \proglang{R} package written in \proglang{C++} that provides Bayesian methods for change point analysis. The package offers two key contributions: (1) a function for detecting change points on time series and epidemic diffusions, which is not available in other \proglang{R} packages, and (2) the implementation of a novel method to cluster time-dependent data sharing common change points. The \proglang{R} interface makes \pkg{BayesChange} accessible to non-advanced users, while the underlying \proglang{C++} implementation ensures computational efficiency. Future developments of \pkg{BayesChange} may include additional kernels for both change points detection and clustering methods. 


\section*{Computational details}

The results in this paper were obtained using \proglang{R}~4.4.3 with the \pkg{BayesChange}~2.1.2 package on a macOS 15.6.1 machine with chip Apple M3, and the dependencies of the following packages: \pkg{Rcpp}~1.1.0, \pkg{RcppArmadillo}~15.0.22, \pkg{RcppGSL}~0.3.13, \pkg{salso}~0.3.57, \pkg{dplyr}~1.1.4, \pkg{tidyr}~1.3.1, \pkg{ggplot2}~4.0.0, and \pkg{ggpubr}~0.6.2. \proglang{R} itself and all packages used are available from the Comprehensive \proglang{R} Archive Network (CRAN) at \url{https://CRAN.R-project.org/}.

\section*{Acknowledgments}

The authors are thankful to Datalab - Bicocca Data Science Lab for providing computational resources to this project. 

\bibliography{references.bib}

\newpage
\appendix

\begin{appendix}

\section{Algorithms} \label{app:algorithms}
Here we show the scheme of the algorithms implemented in \pkg{BayesChange} for change point detection and clustering. Algorithm~\ref{algo:detect_cp} shows the procedure for detecting change points as presented in Section~\ref{subsec:detect_cp}, this is included in \proglang{C++} functions \fct{detect\_cp\_uni}, \fct{detect\_cp\_multi} and \fct{detect\_cp\_epi}. Algorithm~\ref{algo:clust_cp} is the procedure for clustering time series and epidemic diffusions with common change points as detailed in Section~\ref{subsec:clust_data}, and implemented in \proglang{C++} functions \fct{clust\_cp\_uni}, \fct{clust\_cp\_multi} and \fct{clust\_cp\_epi}.
\begin{algorithm}[h]
	\setstretch{0.85}
	\DontPrintSemicolon
	\textbf{input}{ a starting order $\rho_i^{(0)} = \{A_{i1}^{(0)}, \dots, A_{im_i}^{(0)}\}$ of $\bm{y}_i = \{y_{i1}, \dots , y_{iT} \}$, the number $M > 0$ of MCMC iterations and $q \in (0,1)$.}
	
	\For{$m=1,\dots, M$}  { 
		
		\begin{enumerate}
			\item[a)] \textbf{set} $\rho_i^{(N)} = \rho_i^{(m-1)}$, with $\{A_{i1}^{(N)}, \dots, A_{im_i}^{(N)}\}$ denoting the blocks of $\rho_i^{(N)}$. 
			\item[b)] \textbf{with probability $q$} perform a \textbf{split}: 

                \begin{itemize}
                    \item randomly choose one block in $\{A_{i1}^{(N)}, \dots, A_{im_i}^{(N)}\}$;
                    \item sample two consecutive observations and split the chosen block in two new groups and define $\rho_i^{(N)} = \{A_{i1}^{(N)}, \dots, A_{im_i + 1}^{(N)}\}$;
                    \item evaluate the proposal, if accepted $\rho_i^{(m)} = \rho_i^{(N)}$, otherwise $\rho_i^{(m)} = \rho_i^{(m-1)}$.
                \end{itemize}

                \textbf{otherwise} perform a \textbf{merge}: 

                \begin{itemize}
                    \item randomly choose one block in $\{A_{i1}^{(N)}, \dots, A_{im_i}^{(N)}\}$;
                    \item merge the block with the consecutive one and define $\rho_i^{(N)} = \{A_{i1}^{(N)}, \dots, A_{im_i - 1}^{(N)}\}$;
                    \item evaluate the proposal, if accepted $\rho_i^{(m)} = \rho_i^{(N)}$, otherwise $\rho_i^{(m)} = \rho_i^{(m-1)}$.
                \end{itemize}

                \item[c)] \textbf{If $m_i$ > 1} perform a \textbf{shuffle}:
                \begin{itemize}
                    \item set $\rho_i^{(N)} = \rho_i^{(m)}$
                    \item randomly choose one block $A_{ij}^{(N)}$ in $\rho_i^{(N)}$;
                    \item rearrange randomly the observations of $A_{ij}^{(N)}$ and $A_{ij+1}^{(N)}$;
                    \item evaluate the proposal, if accepted $\rho_i^{(m)} = \rho_i^{(N)}$, otherwise $\rho_i^{(m)} = \rho_i^{(m)}$.
                \end{itemize}
			
			\item[d)] \textbf{update} parameters $\sigma$, $\delta$ and $\phi$. 
		\end{enumerate}
		
	}
	\textbf{end}
	\caption{\label{algo:detect_cp} Split-merge algorithm in to detect change points in \fct{detect\_cp}.}
\end{algorithm}

\begin{algorithm}[h] 
	\setstretch{0.85}
	\DontPrintSemicolon
	\textbf{input}{ a partition $\lambda^{(0)} = \{B_1^{(0)}, \dots, B_k^{(0)}\}$ of $\mathcal{Y} = \{\bm{y}_1, \dots, \bm{y}_n\}$, initial values for the unique latent orders $\mathcal R^{*(0)}$} and the number $M > 0$ of MCMC iterations. 
	
	\For{$m=1,\dots, M$}  { 
		
		\begin{enumerate}
			\item[a)] \textbf{set} $\lambda^{(N)} = \lambda^{(m-1)}$, with $\{B_1^{(N)}, \dots, B_k^{(N)}\}$ denoting the blocks \\ of $\lambda^{(N)}$, and $\mathcal R^{*(N)} = \mathcal R^{*(0)}$. 
			\item[b)] \textbf{sample} $i,\ell \in \{1,\dots,n\}$ such that $i \neq \ell$.
			\begin{itemize}
				\item[] \textbf{if} both $i$ and $\ell$ belong to the same block $B_s^{(m-1)}$, perform a split:
				\begin{enumerate}
					\item[i)] assign $i$ to $B_s^{(N)}$ and $\ell$ to a new block $B_{k+1}^{(N)}$;
					\item[ii)] assign randomly each values of $B_s^{(m-1)}$ to $B_s^{(N)}$ or $B_{k+1}^{(N)}$;
					\item[iii)] sample the two distinct unique values of the latent orders in $\mathcal R^{*(N)}$\\ associated with the observations whose indices belong to \\ $B_s^{(N)}$ or $B_{k+1}^{(N)}$ from~(\ref{eq:proposal}).
				\end{enumerate}
				\item[] \textbf{else if} $i$ and $\ell$ belong to different blocks, with $i \in B_s^{(m-1)}$ and\\ $\ell \in B_w^{(m-1)}$, perform a merge:
				\begin{enumerate}
					\item[i)] assign all the indices in $B_w^{(m-1)}$ to $B_s^{(m-1)}$ and destroy $B_w^{(m-1)}$; 
					\item[ii)] sample the  unique value of the latent orders in $\mathcal R^{*(N)}$ associated with \\ all the observations whose indices are in $B_s^{(m-1)}$ from~(\ref{eq:proposal}).
				\end{enumerate}
			\end{itemize}
			
			\item[c)] \textbf{perform} a Metropolis--Hastings step to accept the proposed values: 
                    \begin{itemize}
                        \item[$\rightarrow$] if accepted, set $(\lambda^{(m)}, \mathcal R^{*(m)}) = (\lambda^{(N)}, \mathcal R^{*(N)})$,
                        \item[$\rightarrow$] otherwise, set $(\lambda^{(m)}, \mathcal R^{*(m)}) = (\lambda^{(m-1)}, \mathcal R^{*(m-1)})$.
                    \end{itemize}
                    
                \item[d)] \textbf{update} the unique values $\rho_1^{*(m)}, \dots, \rho_k^{*(m)}$ in $\mathcal R^{*(m)}$.
			
		\end{enumerate}
		
	}
	\textbf{end}
	\caption{Split and merge algorithm that updates $\lambda$ in \fct{clust\_cp}.}
    \label{algo:clust_cp} 
\end{algorithm} 

\section{Code} \label{app:code}

The following is the code for generating synthetic univariate time series for the clustering application in Section~\ref{sec:clust_data_app}.  

\begin{CodeChunk}
\begin{CodeInput}
R> data <- matrix(NA, nrow = 5, ncol = 200)
R> data[1, 1] <- rnorm(n = 1, mean = 0, sd = 0.100)
R> data[2, 1] <- rnorm(n = 1, mean = 0, sd = 0.125)
R> data[3, 1] <- rnorm(n = 1, mean = 0, sd = 0.175)
R> for(i in 2:50){
R>  data[1, i] <- 0.1 * data[1, i-1] + (1 - 0.1) * 0 + 
R>      rnorm(1, mean = 0, sd = (1 - 0.1^2) * 0.100)
R>  data[2, i] <- 0.1 * data[2, i-1] + (1 - 0.1) * 0 + 
R>      rnorm(1, mean = 0, sd = (1 - 0.1^2) * 0.125)
R>  data[3, i] <- 0.1 * data[3, i-1] + (1 - 0.1) * 0 + 
R>      rnorm(1, mean = 0, sd = (1 - 0.1^2) * 0.110)
R> }
R> data[1, 51] <- rnorm(1, mean = 1, sd = 0.230)
R> data[2, 51] <- rnorm(1, mean = 1, sd = 0.225)
R> data[3, 51] <- rnorm(1, mean = 1, sd = 0.240)
R> for(i in 52:150){
R>  data[1, i] <- 0.1 * data[1, i-1] + (1 - 0.1) * 1 + 
R>      rnorm(1, mean = 0, sd = (1 - 0.1^2) * 0.230)
R>  data[2, i] <- 0.1 * data[2, i-1] + (1 - 0.1) * 1 + 
R>      rnorm(1, mean = 0, sd = (1 - 0.1^2) * 0.225)
R>  data[3, i] <- 0.1 * data[3, i-1] + (1 - 0.1) * 1 + 
R>      rnorm(1, mean = 0, sd = (1 - 0.1^2) * 0.240)
R>}
R> data[1, 151] <- rnorm(1, mean = 0.5, sd = 0.225)
R> data[2, 151] <- rnorm(1, mean = 0.5, sd = 0.235)
R> data[3, 151] <- rnorm(1, mean = 0.5, sd = 0.100)
R> for(i in 152:200){
R>  data[1, i] <- 0.1 * data[1, i-1] + (1 - 0.1) * 0 + 
R>      rnorm(1, mean = 0, sd = (1 - 0.1^2) * 0.225)
R>  data[2, i] <- 0.1 * data[2, i-1] + (1 - 0.1) * 0 + 
R>      rnorm(1, mean = 0, sd = (1 - 0.1^2) * 0.235)
R>  data[3, i] <- 0.1 * data[3, i-1] + (1 - 0.1) * 0 + 
R>      rnorm(1, mean = 0, sd = (1 - 0.1^2) * 0.100)
R> }
R> data[4, 1] <- rnorm(1, mean = 0, sd = 0.135)
R> data[5, 1] <- rnorm(1, mean = 0, sd = 0.155)
R> for(i in 2:25){
R>  data[4, i] <- 0.1 * data[4,i-1] + (1 - 0.1) * 0 + 
R>      rnorm(1, mean = 0, sd = (1 - 0.1^2) * 0.135)
R>  data[5, i] <- 0.1 * data[5,i-1] + (1 - 0.1) * 0 + 
R>      rnorm(1, mean = 0, sd = (1 - 0.1^2) * 0.155)
R> }
R> data[4, 26] <- rnorm(1, mean = 1, sd = 0.165)
R> data[5, 26] <- rnorm(1, mean = 1, sd = 0.185)
R> for(i in 27:200){
R>  data[4, i] <- 0.1 * data[4, i-1] + (1 - 0.1) * 1 + 
R>      rnorm(1, mean = 0, sd = (1 - 0.1^2) * 0.165)
R>  data[5, i] <- 0.1 * data[5, i-1] + (1 - 0.1) * 1 + 
R>      rnorm(1, mean = 0, sd = (1 - 0.1^2) * 0.185)
R> }
\end{CodeInput}
\end{CodeChunk}

The following is the code for generating synthetic multivariate time series for the clustering application in Section~\ref{sec:clust_data_app}.  

\begin{CodeChunk}
\begin{CodeInput}
R> data <-  array(data = NA, dim = c(2, 200, 5))
R> data[1:2,1,1] <- rnorm(1 ,mean = 0, sd = 0.100)
R> data[1:2,1,2] <- rnorm(1, mean = 0, sd = 0.125)
R> data[1:2,1,3] <- rnorm(1, mean = 0, sd = 0.175)
R> for(i in 2:50){
R>  data[1, i, 1] <- 0.1 * data[1, i-1, 1] + (1 - 0.1) * 0 + 
R>      rnorm(1, mean = 0, sd = (1 - 0.1^2) * 0.100)
R>  data[2, i, 1] <- 0.1 * data[2, i-1, 1] + (1 - 0.1) * 0 + 
R>      rnorm(1, mean = 0, sd = (1 - 0.1^2) * 0.100)
R>  data[1, i, 2] <- 0.1 * data[1, i-1, 2] + (1 - 0.1) * 0 + 
R>      rnorm(1, mean = 0, sd = (1 - 0.1^2) * 0.125)
R>  data[2, i, 2] <- 0.1 * data[2, i-1, 2] + (1 - 0.1) * 0 + 
R>      rnorm(1, mean = 0, sd = (1 - 0.1^2) * 0.125)
R>  data[1, i, 3] <- 0.1 * data[1, i-1, 3] + (1 - 0.1) * 0 + 
R>      rnorm(1 ,mean = 0, sd = (1 - 0.1^2) * 0.110)
R>  data[2, i, 3] <- 0.1 * data[2, i-1, 3] + (1 - 0.1) * 0 + 
R>      rnorm(1, mean = 0, sd = (1 - 0.1^2) * 0.110)
R> }
R> data[1, 51, 1] <- data[2, 51, 1] <- rnorm(1, mean = 1, sd = 0.230)
R> data[1, 51, 2] <- data[2, 51, 2] <- rnorm(1, mean = 1, sd = 0.225)
R> data[1, 51, 3] <- data[2, 51, 3] <- rnorm(1, mean = 1, sd = 0.240)
R> for(i in 52:150){
R>  data[1, i, 1] <- 0.1 * data[1, i-1, 1] + (1 - 0.1) * 1 + 
R>      rnorm(1, mean = 0, sd = (1 - 0.1^2) * 0.230)
R>  data[2, i, 1] <- 0.1 * data[2, i-1, 1] + (1 - 0.1) * 1 + 
R>      rnorm(1, mean = 0, sd = (1 - 0.1^2) * 0.230)
R>  data[1, i, 2] <- 0.1 * data[1, i-1, 2] + (1 - 0.1) * 1 + 
R>      rnorm(1, mean = 0, sd = (1 - 0.1^2) * 0.225)
R>  data[2, i, 2] <- 0.1 * data[2, i-1, 2] + (1 - 0.1) * 1 + 
R>      rnorm(1, mean = 0, sd = (1 - 0.1^2) * 0.225)
R>  data[1, i, 3] <- 0.1 * data[1, i-1, 3] + (1 - 0.1) * 1 + 
R>      rnorm(1, mean = 0, sd = (1 - 0.1^2) * 0.240)
R>  data[2, i, 3] <- 0.1 * data[2, i-1, 3] + (1 - 0.1) * 1 + 
R>      rnorm(1, mean = 0, sd = (1 - 0.1^2) * 0.240)
R> }
R> data[1:2, 151, 1] <- rnorm(2, mean = 0.5, sd = 0.225)
R> data[1:2, 151, 2] <- rnorm(2, mean = 0.5, sd = 0.235)
R> data[1:2, 151, 3] <- rnorm(2, mean = 0.5, sd = 0.100)
R> for(i in 152:200){
R>  data[1, i, 1] <- 0.1 * data[1, i-1, 1] + (1 - 0.1) * 0.5 + 
R>      rnorm(1, mean = 0, sd = (1 - 0.1^2) * 0.225)
R>  data[2, i, 1] <- 0.1 * data[2, i-1, 1] + (1 - 0.1) * 0.5 + 
R>      rnorm(1, mean = 0, sd = (1 - 0.1^2) * 0.225)
R>  data[1, i, 2] <- 0.1 * data[1, i-1, 2] + (1 - 0.1) * 0.5 + 
R>      rnorm(1, mean = 0, sd = (1 - 0.1^2) * 0.235)
R>  data[2, i, 2] <- 0.1 * data[2, i-1, 2] + (1 - 0.1) * 0.5 + 
R>      rnorm(1, mean = 0, sd = (1 - 0.1^2) * 0.235)
R>  data[1, i, 3] <- 0.1 * data[1, i-1, 3] + (1 - 0.1) * 0.5 + 
R>      rnorm(1, mean = 0, sd = (1 - 0.1^2) * 0.100)
R>  data[2, i, 3] <- 0.1 * data[2, i-1, 3] + (1 - 0.1) * 0.5 + 
R>      rnorm(1, mean = 0, sd = (1 - 0.1^2) * 0.100)
R> }
R> data[1:2,1,4] <- rnorm(1, mean = 1, sd = 0.135)
R> data[1:2,1,5] <- rnorm(1, mean = 1, sd = 0.155)
R> for(i in 2:25){
R>  data[1, i, 4] <- 0.1 * data[1, i-1, 4] + (1 - 0.1) * 1 + 
R>      rnorm(1, mean = 0, sd = (1 - 0.1^2) * 0.135)
R>  data[2, i, 4] <- 0.1 * data[2, i-1, 4] + (1 - 0.1) * 1 + 
R>      rnorm(1, mean = 0, sd = (1 - 0.1^2) * 0.135)
R>  data[1, i, 5] <- 0.1 * data[1, i-1, 5] + (1 - 0.1) * 1 + 
R>      rnorm(1, mean = 0, sd = (1 - 0.1^2) * 0.155)
R>  data[2, i, 5] <- 0.1 * data[2, i-1, 5] + (1 - 0.1) * 1 + 
R>      rnorm(1, mean = 0, sd = (1 - 0.1^2) * 0.155)
R> }
R> data[1:2, 26, 4] <- rnorm(n = 1, mean = 0.5, sd = 0.165)
R> data[1:2, 26, 5] <- rnorm(n = 1, mean = 0.5, sd = 0.185)
R> for(i in 27:200){
R>  data[1, i, 4] <- 0.1 * data[1, i-1, 4] + (1 - 0.1) * 0.5 + 
R>      rnorm(1, mean = 0, sd = (1 - 0.1^2) * 0.165)
R>  data[2, i, 4] <- 0.1 * data[2, i-1, 4] + (1 - 0.1) * 0.5 + 
R>      rnorm(1, mean = 0, sd = (1 - 0.1^2) * 0.165)
R>  data[1, i, 5] <- 0.1 * data[1, i-1, 5] + (1 - 0.1) * 0.5 + 
R>      rnorm(1, mean = 0, sd = (1 - 0.1^2) * 0.185)
R>  data[2, i, 5] <- 0.1 * data[2, i-1, 5] + (1 - 0.1) * 0.5 + 
R>      rnorm(1, mean = 0, sd = (1 - 0.1^2) * 0.185)
R> }
\end{CodeInput}    
\end{CodeChunk}

\end{appendix}

\end{document}